\documentclass[useAMS,usenatbib]{mn2e}
\usepackage{epsfig}

\def\ej{{\rm ej}}
\def\rem{{\rm rem}}
\def\in{{\rm in}}
\def\br{{\rm br}}

\def\ph{{\rm ph}}
\def\p{{\rm peak}}
\def\iso{{\rm iso}}
\def\es{{\rm es}}
\def\x{{\rm X}}
\def\l{{\rm light}}

\def\sw{{\it Swift}}
\def\hst{{\it HST}}

\def\obs{{\rm obs}}

\title[Shock Breakout and GRB 060218/SN 2006aj]
{Shock Breakout in Type Ibc Supernovae and Application to GRB~060218/SN~2006aj}
\author[Li-Xin Li]{Li-Xin Li\thanks{E-mail: lxl@mpa-garching.mpg.de}\\
Max-Planck-Institut f\"ur Astrophysik, 85741 Garching, Germany}

\begin{document}

\date{Accepted 2006 November 13. Received 2006 November 2; in original form
2006 May 16}

\pagerange{\pageref{firstpage}--\pageref{lastpage}} \pubyear{2006}

\maketitle

\label{firstpage}

\begin{abstract}
Recently, a soft black-body component was observed in the early X-ray
afterglow of GRB 060218, which was interpreted as shock breakout from the 
thick wind of the progenitor Wolf-Rayet (WR) star of the underlying Type Ic 
SN 2006aj. In this paper we present a simple model for computing the 
characteristic quantities (including energy, temperature, and time-duration) 
for the transient event from the shock breakout in Type Ibc supernovae 
produced by the core-collapse of WR stars surrounded by dense winds. 
In contrast to the case of a star without a strong wind, the shock breakout 
occurs in the wind region rather than inside the star, caused by the large 
optical depth in the wind. We find that, for the case of a WR star 
with a dense wind, the total energy of the radiation generated by the supernova
shock breakout is larger than that in the case of the same star without a 
wind by a factor $> 10$. The temperature can be either hotter or cooler, 
depending on the wind parameters. The time-duration is larger caused by the 
increase in the effective radius of the star due to the presence of a thick 
wind. Then, we apply the model to GRB 060218/SN 2006aj. We show that, to 
explain both the temperature
and the total energy of the black-body component observed in GRB 060218
by the shock breakout, the progenitor WR star has to have an unrealistically 
large core radius (the radius at optical depth of 20), larger than $100 
R_\odot$. In spite of this disappointing result, our model is expected to have 
important applications to the observations on Type Ibc supernovae in which 
the detection of shock breakout will provide important clues to the 
progenitors of SNe Ibc.
\end{abstract}


\section{Introduction}
\label{intro}

Since the first detection of the afterglows \citep{cos97,par97,fra97} and the
host galaxies \citep{blo98,blo99,fru99b} of gamma-ray bursts (GRBs), by now 
it has well been established that long-duration GRBs are cosmological events
occurring in star-forming galaxies 
\citep[and references therein]{pac98a,fru99a,ber01,fra02,chr04,sol05,fru06}, 
and are most likely produced by the core-collapse of massive stars 
(Woosley, Heger \& Weaver 2002; Piran 2004; Zhang \& M\'esz\'aros 2004;
Woosley \& Heger 2006a, and references therein). This scenario has received 
strong support from the cumulative evidence that some, if 
not all, long-duration GRBs are associated with supernovae (SNe), either 
from direct observations of supernova features in the spectra of GRB 
afterglows, or from indirect observations of rebrightening and/or flattening 
(called ``red bumps'') in GRB afterglows which are interpreted as the 
emergence of the underlying supernova lightcurves 
\citep[and references therein]{del06,woo06x,woo06}. The discovery of the 
connection between GRBs and supernovae has been one of the most exciting 
developments in the fields of GRBs and supernovae in the past decade.

Interestingly, all the supernovae that have been spectroscopically confirmed
to be associated with GRBs, including SN 1998bw/GRB 980425 \citep{gal98}, 
SN 2003dh/GRB 030329 \citep{hjo03,sta03}, SN 2003lw/GRB 031203 (Malesani et 
al. 2004; Sazonov, Lutovinov \& Sunyaev 2004), and the most recent one, 
SN 2006aj/GRB 060218 
\citep{mas06,mod06,cam06,sol06,pia06,mir06,cob06}, are Type Ic having
no detectable hydrogen and helium lines. However, the supernovae that are 
associated with GRBs also remarkably differ from ordinary Type Ibc supernovae:
they have extremely smooth and featureless spectra indicating very large 
expansion velocity, are much more energetic (i.e., involving much larger 
explosion energy), and eject significantly larger amount of nickels 
\citep{ham04,del06,woo06}, except SN 2006aj/GRB 060218 which is somewhat 
closer to normal SNe Ibc (see below; Mazzali et al. 2006). For these reasons, 
they are often called ``hypernovae'' to be distinguished from normal 
supernovae \citep{iwa98,pac98a,pac98b}. A correlation between the peak 
spectral energy of GRBs and the peak bolometric luminosity of the underlying 
supernovae are found by \citet{li06}, based on the multi-wavelength 
observations on the above four pairs of GRBs-SNe. 

The discovery of GRB-SN connection has provided us with important clues to 
the progenitors of GRBs, since it is broadly believed that Type Ibc 
supernovae are produced by the core-collapse of Wolf-Rayet (WR) stars who 
have lost their hydrogen (possibly also helium) envelopes due to strong 
stellar winds or interaction with companions 
\citep[and references therein]{sma02,woo02,fil04,woo06a}. In fact, for 
several GRBs, observations with high quality optical spectra have identified
the presence of highly ionized lines with high relative velocities most
likely coming from shells or clumps of material from WR stars, supporting
WR stars as the GRB progenitors 
\citep[see, however, Hammer et al. 2006]{mir03,sch03,klo04,che06}.

A systematic study on the GRB afterglows carried out by \citet{zeh04} 
suggested that all long-duration GRBs are associated with supernovae. However, 
it appears that only a small fraction of Type Ic supernovae are able to 
produce GRBs, since the rate of GRBs and hypernovae are several orders of 
magnitude lower than the rate of core-collapse supernovae \citep{pod04}. 
Although both long-duration GRBs and core-collapse supernovae are found in 
star-forming galaxies, their location in the hosts and the morphology and 
luminosities of their host galaxies are significantly different as most 
clearly revealed by the recent study of \citet{fru06} with {\it Hubble Space 
Telescope} (\hst) imaging. The core-collapse supernovae trace the blue-light 
of their hosts that are approximately equally divided between spiral and 
irregular galaxies, while long GRBs are 
far more concentrated on the brightest regions of faint and irregular 
galaxies. \citet{fru06} argued that their results may be best understood if 
GRBs are formed from the collapse of extremely massive and low-metallicity 
stars.

The preference of long-duration GRBs to low-metallicity galaxies
\citep{fyn03,hjo03a,lef03,sol05,fru06} has been strengthened by the recent 
paper of \citet{sta06}, in which a strong anti-correlation between the 
isotropic energy of five nearby SN-connected GRBs and the oxygen abundance 
in their host galaxies was found, which was used to argue that the life in 
the Milky Way is protected away from GRBs by metals. \citet{sta06} have 
suggested that long-duration GRBs do not trace star formation, but trace the 
metallicity.

The discovery of GRB 060218 and its association with SN 2006aj by \sw\, has 
shed more light on the GRB-SN connection as well as on the nature of GRBs. 
GRB~060218 has a cosmological redshift $z = 0.0335$ corresponding to a 
luminosity distance of $147 {\rm Mpc}$ ($\Omega_m = 0.3$, $\Omega_\Lambda =
0.7$, and $H_0 = 70 ~{\rm km~s}^{-1} {\rm Mpc}^{-1}$), which makes it the 
second nearest GRB among those having determined redshifts (about four times 
the distance of GRB 980425 at $z=0.0085$; Campana et al. 2006; Pian et al.
2006; Sollerman et al. 2006). GRB 060218 is very unusual in several aspects. 
It has an extremely long duration, about $2,100$ s. Its spectrum is very soft, 
with a photon index $2.5\pm 0.1$ and peak energy $E_\p = 4.9_{-0.3}^{+0.4}\,
{\rm keV}$ in the GRB frame. The isotropic equivalent energy is $E_\iso = 
(6.2\pm 0.3)\times 10^{49}$ ergs extrapolated to the $1$--$10,000$ keV in 
the rest frame energy band \citep{cam06}, which is at least 100 times fainter 
than normal cosmological GRBs but among a population of under-energetic 
GRBs \citep{saz04,lia06}. 

Although the supernova associated with GRB 060218, i.e. SN 2006aj, is broadly 
similar to those previously discovered GRB-connected supernovae, it also shows 
some remarkable unusual features \citep{pia06,sol06,maz06a}. Among the four 
GRB-connected supernovae mentioned above, SN 2006aj is the faintest one, 
although still brighter than normal Type Ibc supernovae. Its lightcurve rises 
more rapidly, and its expansion velocity indicated by the spectrum is 
intermediate between that of other GRB-connected supernovae and that of 
normal SNe Ibc. Modeling of the spectra and the lightcurve of SN 2006aj 
reveals that SN 2006aj 
is much less energetic compared to other GRB-connected supernovae: it had an 
explosion energy $E_\in \approx 2\times 10^{51}$ ergs, ejected a mass 
$M_\ej \approx 2 M_\odot$, compared to $E_\in \sim 3$--$6\times 10^{52}$ 
ergs, and $M_\ej\sim 10 M_\odot$ of the others \citep{maz06a}. This suggests 
that SN 2006aj is closer to normal Type Ibc supernovae than to the other 
GRB-connected supernovae, and there does not exist a clear gap between 
hypernovae and normal Type Ibc supernovae \citep{li06}.

The X-ray afterglow observation by the X-Ray Telescope (XRT) on board \sw\,
on GRB 060218 started 159~s after the burst trigger. A very interesting 
feature in the early X-ray afterglow is that it contains a soft black-body 
component which has a temperature about $0.17$ keV and comprises about 
$20\%$ of the total X-ray flux in the $0.3$--$10$ keV range, lasting from 
159~s up to $\sim 10,000$~s. The black-body component was not detected in 
later XRT observations \citep{cam06}. The total energy contained in the
black-body component, as estimated by Campana (private communication), is 
$\approx 10^{49}$ ergs. \citet{cam06} interpreted it as supernova shock 
breakout from a dense wind surrounding the progenitor WR star of the 
supernova.

\citet{but06} conducted an analysis on the early X-ray afterglows of a 
sample ($>70$) of GRBs observed by the XRT/\sw. He found that although most of
the afterglow spectra can be fitted with a pure power law with extinction, a
small fraction of them show appreciable soft thermal components at $5$--$10\%$
level. His reanalysis on GRB 060218 showed that the black-body component
contains energy as much as $2.3\times 10^{50}$ ergs and has a duration 
$\approx 300$ s. According to Butler's analysis, the soft black-body component 
even dominates the flux after $\sim 1,000$~s from the burst trigger.

Flashes from shock breakout in core-collapsed supernovae were first predicted 
by \citet{col68} almost forty years ago, originally proposed for GRBs that had 
not been discovered yet. However, they have not been unambiguously detected in 
supernova observations yet \citep{cal04}. This is mainly due to the transient 
nature of the event. It is generally expected that the flash from shock 
breakout precedes the supernova, is much brighter and harder than the 
supernova radiation but has a very short time-duration.

According to the general theory of core-collapsed supernova explosion, the
liberation of explosive energy in the interior of a progenitor star generates
a shock wave. The shock wave propagates outward. However, the external 
appearance of the star remains unaltered, until the shock wave reaches a 
point (the shock breakout point) near the stellar surface where the diffusion 
velocity of photons begins to exceed the shock velocity. The postshock 
radiation can then leak out in a burst of ionizing radiation, producing a 
brilliant flash in the UV/X-ray band \citep[for a comprehensive review see 
Matzner \& McKee 1999]{kle78,che79,ims89}.

For the famous Type II SN 1987A, theoretical calculations have shown that 
the shock emergence from the surface of the progenitor (Sk~1, a blue 
supergiant) would have produced a radiation of $\sim 10^{47}$ ergs in the 
extreme UV to soft X-ray band, lasting 1--3 minutes 
\citep{ims88,ims89,ens92,bli00}. 
In fact, in the observed bolometric lightcurve of SN 1987A, there was a fast 
initial decline phase which could be the tail of the lightcurve produced by 
the shock breakout \citep{ims89}. If the shock breakout interpretation of 
the soft black-body component in GRB 060218 is confirmed, it would have 
important impact on the theories of both GRBs and supernovae. The case of 
GRB 060218/SN 2006aj would also be the first unambiguous detection of a shock
breakout event from supernovae.

Although the propagation of a strong shock in a supernova and the appearance 
of shock emergence (shock breakout) have been intensively studied both
analytically and numerically (Klein \& Chevalier 1978; Imshennik \& Nad\"ezhin
1988, 1989; Ensman \& Burrows 1992; Blinnikov et al. 1998, 2000, 2002; 
Matzner \& McKee 1999; Tan, Matzner \& McKee 2001), in the 
situation of supernovae produced from stars with dense stellar winds they have 
not been fully explored yet. If the stellar wind of the progenitor is very 
optically thick---which is indeed the case for Type Ibc supernovae whose 
progenitors are believed to be WR stars---the shock breakout will occur in 
the wind region after the shock passes through the surface of the 
star, instead of in the region inside the star. Since a stellar wind has a 
mass density profile very different from that of a star, the model 
that has been developed for the shock emergence in supernovae with progenitors 
without stellar winds cannot be directly applied to the case of progenitors 
with dense stellar winds. 

In this paper, we present a simple model for semi-analytically computing 
the propagation of a strong shock in a dense stellar wind, and estimating 
the characteristic quantities for the transient event from the shock breakout 
in SNe Ibc. The model is obtained by an extension of the existing model for 
the shock propagation and breakout in supernovae produced by the core-collapse 
of stars without dense 
stellar winds. Then, we apply the model to SN 2006aj and examine if the 
soft black-body component in the early X-ray afterglow of GRB 060218 can
be interpreted by the supernova shock breakout.

The paper is organized as follows. In Sec.~\ref{wr}, we describe a simple
but general model for the mass density and velocity profile for the wind 
around a WR star, and calculate the optical depth in the wind. In 
Sec.~\ref{shock}, we model the propagation of a supernova shock wave in a 
stellar wind, taking into account the relativistic effects. In 
Sec.~\ref{energy}, we analyze the evolution of the shock front, and 
the radiation energy contained in it. In Sec.~\ref{emergence}, we present
a procedure for calculating the quantities characterizing the transient event
arising from the shock breakout, including the released energy, the 
temperature, the time-duration, and the momentum of the shock front at the
time of shock breakout. In Sec.~\ref{result}, we present our numerical 
results. In Sec.~\ref{application}, we apply our model to 
GRB 060218/SN 2006aj. In Sec.~\ref{sum}, we summarize our results and draw 
our conclusions.

Appendix~\ref{b1} is devoted to the formulae for computing the optical depth 
of a wind in the framework of the standard stellar wind model. 
Appendix~\ref{star} lists the formulae for computing the characteristic
quantities for supernova shock breakout from a star without winds, in the 
trans-relativistic regime. Appendix~\ref{correlation} presents a correlation
in the WR star parameters.

\section{Mass Density Profile of the Wind of a Wolf-Rayet Star and the Optical
Depth in the Wind}
\label{wr}

Wolf-Rayet stars are very luminous, hot, and massive stars that are nearly 
at the end of their stellar lives. Based on their spectra, WR stars are often 
classified as WN stars (nitrogen dominant) and WC stars (carbon dominant).
WR stars are characterized by extremely dense stellar winds, losing mass at
a rate of $10^{-6}$--$10^{-4} M_\odot$ yr$^{-1}$ with a wind terminal velocity 
of $700$--$3,000$ km s$^{-1}$. The mass lost from WR stars by stellar 
winds is so enormous that most (if not all) of hydrogens of WR stars have 
been lost. This is a main reason for the general belief that WR stars are the 
progenitors of Type Ibc supernovae. 

Some basic relations in the physical parameters of WR stars can be found
in \citet{lan89}, \citet{sch92}, and \citet{nug00}.

The wind of a WR star is usually extremely dense. This is characterized by
the fact that, for the majority of WR stars, the ratio of the momentum of
the wind ($\dot{M} v_\infty$, where $\dot{M}$ is the mass-loss rate, 
$v_\infty$ is the terminal velocity of the wind) to the momentum of 
radiation ($L/c$, where $L$ is the luminosity of the star, $c$ is the speed
of light) is much larger than unity, indicating that on average each photon 
leaving the star must be scattered several times and the wind must be 
optically thick. As a result, the photospheric radius ($R_\ph$, the radius 
where the optical depth $\tau_w = 2/3$) often differs from the core radius 
of the star ($R_\star$, the radius where $\tau_w = 20$ by definition) 
by a factor $> 2$.

\begin{figure}
\vspace{2pt}
\includegraphics[angle=0,scale=0.47]{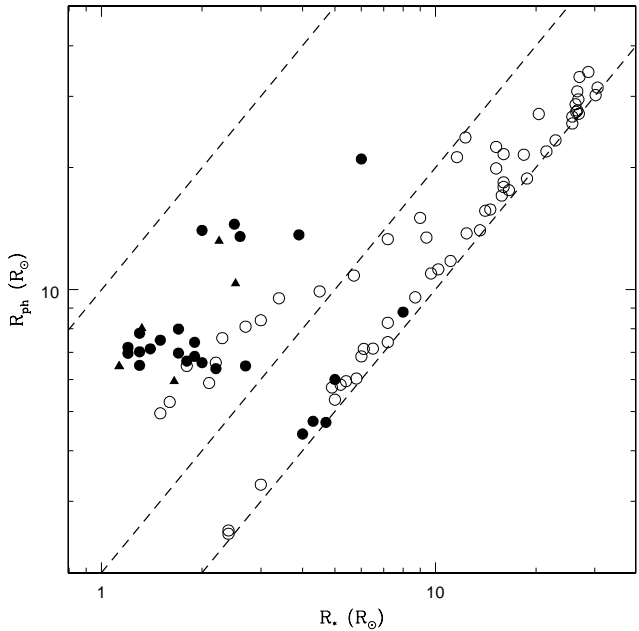}
\caption{Photospheric radius versus stellar core radius, for 86 Galactic WRs 
(filled circles for WC-type, open circles for WN-type) and 6 LMC WRs 
(triangles, WC-type only). The dashed lines show the relation of $R_\ph 
= R_\star$, $R_\ph = 2 R_\star$, and $R_\ph = 10 R_\star$ (upward). 
}
\label{rph}
\end{figure}

In Fig.~\ref{rph}, we plot the photospheric radius against the core radius 
for 86 Galactic WC-type and WN-type stars (Hamann, Koesterke \& Wessolowski
1995; Koesterke \& Hamann 1995) and 6 LMC WC-type stars \citep{gra98}, 
determined with the ``standard model'' of stellar winds. For many WRs, 
especially those of WC-type, we have $R_\ph > 2 R_\star$.

The mass density of a steady and spherically symmetric wind is related to
the mass-loss rate and the wind velocity by
\begin{eqnarray}
	\rho(r) = \frac{\dot{M}}{4\pi r^2 v_r(r)} \;, \label{rho_w}
\end{eqnarray}
where $r$ is the radius from the center of the star, and $v_r$ is the radial
velocity of the wind. We model the velocity of the wind by 
\citep{sch96,ign00,nug02}
\begin{eqnarray}
	v_r(r) = v_\infty \left(1-\frac{\alpha R_\star}{r}\right)^b \;,
	\label{vr}
\end{eqnarray}
where $\alpha<1$ and $b\ge 1$ are free parameters. The presence of $\alpha$ 
in equation~(\ref{vr}) is to ensure that the mass density of the wind is 
regular at the stellar radius $r=R_\star$. 

In the ``standard model'' of stellar winds the value of $b$ is assumed to 
be unity, as in the case of O-stars. However, it has been argued that for WR 
stars $b$ can be significantly larger \citep{rob94,sch97,lep99}. According 
to the calculations of \citet{nug02}, $b$ is typically in the range of 
$4$--$6$. 

The value of $\alpha$ can be determined by the radial velocity of the wind
at $r=R_\star$. If we define $\varepsilon = v_\star/v_\infty$, where
$v_\star \equiv v_r(R_\star)$, then
\begin{eqnarray}
	\alpha = 1-\varepsilon^{1/b} \;.
	\label{alp}
\end{eqnarray}
Typically, $v_\star$ has the order of the sound speed at $R_\star$, and 
$\varepsilon \sim 0.01$ \citep{sch96}.

\begin{figure}
\vspace{2pt}
\includegraphics[angle=0,scale=0.475]{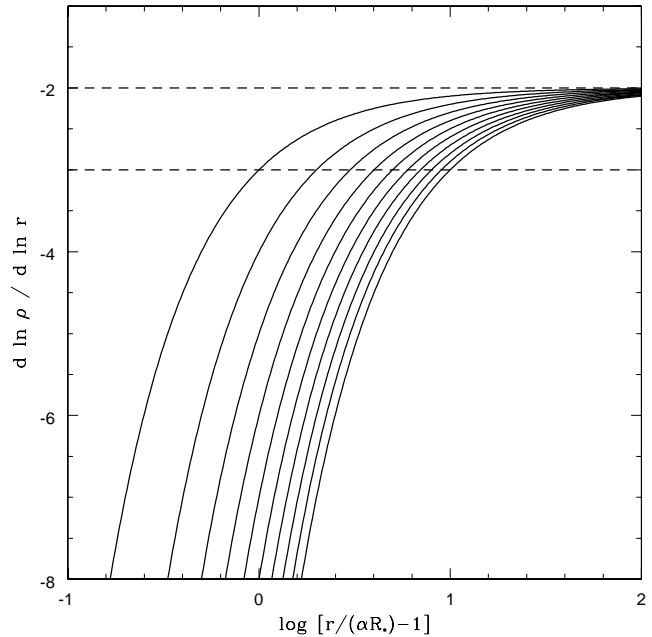}
\caption{The log-slope of the wind density, $s\equiv d\ln\rho/d\ln r = -2 -b
(r/\alpha R_\star -1)^{-1}$, as a function of the radius $r$. As $r\rightarrow
\infty$, $s$ approaches $-2$ (the upper dashed line). For small $r$, $s$ is
significantly smaller than $-2$. A shock wave accelerates only in the region 
of $s<-3$ (below the lower dashed line; see Sec.~\ref{shock}). Left to right: 
$b= 1$--$10$ with $\Delta b= 1$. 
}
\label{slope}
\end{figure}

In the outer wind region, where $r\gg R_\star$ and $v_r\approx v_\infty$, 
the wind density $\rho\sim r^{-2}$. In the region close to the stellar 
surface (i.e., $r\sim R_\star$), the wind density has a much steeper 
log-slope (Fig.~\ref{slope}). As will be seen in Sec.~\ref{shock}, it is the 
very steep mass density profile near the surface of the star that makes it
possible for a shock wave to accelerate in the wind region. We will also see
in Sec.~\ref{result} that shock breakout takes place at a radius not far
from the surface of $r=R_\star$. So we adopt equation~(\ref{vr}) for the
wind velocity profile since its asymptotic form $v_r = v_\infty$ (and $\rho
\propto r^{-2}$) is not accurate for describing the wind velocity (and hence 
the mass density) near $r=R_\star$.

The opacity $\kappa_w$ in the WR wind region is complex and generally a 
function of radius \citep{nug02,gra05}. However, compared to the mass density,
the opacity changes very slowly with radius. For example, at the sonic point
in the wind, we have $d\ln\kappa_w/d\ln r \sim 0.001$--$0.03$ \citep{nug02}, 
while $|d\ln\rho/d\ln r| \ga 2$ always. Hence, to calculate the optical depth
in the wind, we can approximate $\kappa_w$ by a constant although its value is
uncertain at some level. Then, the optical depth in the wind is
\begin{eqnarray}
	\tau_w \equiv \int_r^\infty \kappa_w\rho dr = \frac{A}{(b-1) 
	        \alpha R_\star} \left[\left(1-\frac{\alpha}{y}\right)^{1-b} 
	        -1\right] \;,
	\label{tau}
\end{eqnarray}
where $y\equiv r/R_\star$ and $A\equiv \kappa_w\dot{M}/(4\pi v_\infty)$. 

\begin{figure}
\vspace{2pt}
\includegraphics[angle=0,scale=0.475]{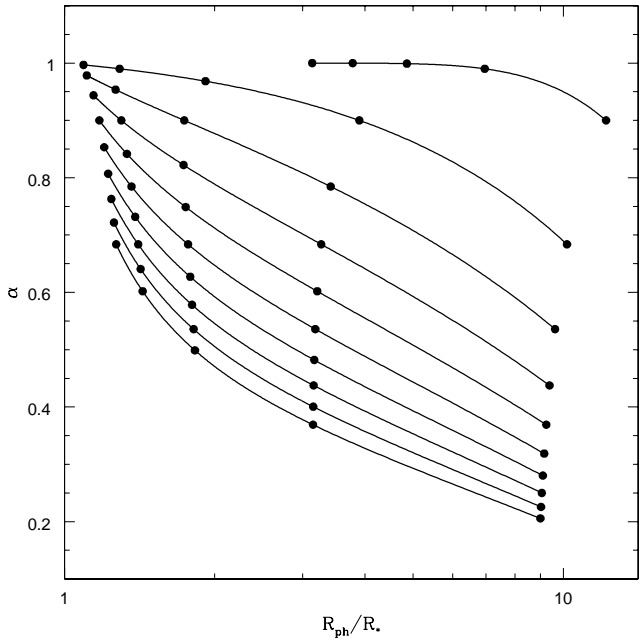}
\caption{The solution for $\alpha$ and $y_\ph = R_\ph/R_\star$, for
$\varepsilon=10^{-5}-10^{-1}$. Top to bottom: $b= 1 - 10$ with $\Delta b
= 1$. The filled circles on each curve label the values of $\varepsilon 
= 10^{-5}$, $10^{-4}$, $10^{-3}$, $10^{-2}$, and $10^{-1}$ from left to 
right.
}
\label{alpha}
\end{figure}

As commonly adopted in the literature, we define the stellar core radius 
$R_\star$ of the WR star to be the radius where $\tau_w = 20$. Then, we can 
rewrite the optical depth as
\begin{eqnarray}
	\tau_w = \tau_0 \left[\left(1-\frac{\alpha}{y}\right)^{1-b} 
	        -1\right] \;,
	\label{tau_r}
\end{eqnarray}
where
\begin{eqnarray}
	\tau_0 \equiv\frac{A}{(b-1)\alpha R_\star} = 
		\frac{20}{(1-\alpha)^{1-b} -1} \;.
	\label{tau0}
\end{eqnarray}

By definition, the boundary of the photosphere is at the photospheric 
radius where $\tau_w = 2/3$. Then, we can solve for $y_\ph \equiv R_\ph/
R_\star$ from equation~(\ref{tau_r})
\begin{eqnarray}
	y_\ph = \alpha \left[1-\left(1+\frac{2}{3\tau_0}
		\right)^{1/(1-b)}\right]^{-1} \;.
	\label{yph}
\end{eqnarray}

For a given $b$, $y_\ph$ is a decreasing function of $\alpha$. As $\alpha
\rightarrow 1$, we have $\tau_0\rightarrow 0$ and $y_\ph\rightarrow 1$. As 
$\alpha\rightarrow 0$, we have $\tau_0\approx 20/[(b-1)\alpha]$ and $y_\ph
\rightarrow 30$. Thus, in general we must have $1<y_\ph <30$. By 
equation~(\ref{alp}), $\alpha$ is a decreasing function of $\varepsilon$.

The above results for the optical depth are valid only for $b>1$. The 
corresponding formulae for $b=1$ (the ``standard model'') are given in 
Appendix~\ref{b1}.

In Fig.~\ref{alpha}, we plot $\alpha$ versus $y_\ph$, solved from 
equations~(\ref{alp}) and (\ref{yph}) (or eq.~\ref{yph1} when $b=1$) for a 
set of discrete values of $b$ (from 1 to 10) and continuous values of 
$\varepsilon$ (from $10^{-5}$ to $10^{-1}$). The value of $y_\ph$ sensitively 
depends on $\varepsilon$. For the same value of $b$, $y_\ph$ drops quickly as 
$\varepsilon$ decreases. When $\varepsilon$ is fixed, $y_\ph$ decreases 
if $b$ increases from $b=1$, very fast for small values of $\varepsilon$.
However, for $b>3$, the effect of the variation in $b$ on the value of
$y_\ph$ is not dramatic.

The opacity $\kappa_w$, and the corresponding optical depth $\tau_w$, are
for the optical photons in the wind and are hence valid for calculating the
mass density profile of the wind before a supernova shock passes through it. As
will be seen in the following sections, the radiation generated by the 
supernova shock wave in the wind of a WR star is in the X-ray band and we 
also need consider the opacity and the optical depth to the X-ray photons for
computing the thickness of the shock front and the emergence of the shock 
wave (Secs.~\ref{energy} and \ref{emergence}). 

The X-ray opacity of a gas strongly depends on the ionization state of the
gas \citep{kro84}. The radiation generated by a supernova shock wave in 
the wind of a WR star has a luminosity $L_\x \ga 10^{45}$ ergs s$^{-1}$ 
(Secs.~\ref{emergence} and \ref{result}), which contains enough photons to
fully ionize the surrounding gas. This fact can be seen from the ionization 
parameter, defined as the ratio of the photon number density to the particle 
number density, $\Xi \equiv L_\x/\left(4\pi c r^2 n_{\rm H} \varepsilon_{\rm 
ph}\right)$, where $\varepsilon_{\rm ph}$ is the energy of photons. Using 
equation~(\ref{rho_w}) (where $v_r\approx v_\infty$) and $n_{\rm H} = \rho/
\mu_{\rm H} m_{\rm H}$, where $m_{\rm H}$ is the mass of proton, and 
$\mu_{\rm H} \approx 2$ is the mean molecular weight per proton, we get
\begin{eqnarray}
	\Xi &\approx& \frac{\mu_{\rm H} m_{\rm H} L_\x v_\infty}{\dot{M}
		\varepsilon_{\rm ph} c} \nonumber \\
		&\approx& 4.4\times 10^6\;\mu^{-1}
		\left(\frac{L_\x}{10^{45}{\rm ergs}\, {\rm s}^{-1}}\right)
		\left(\frac{\varepsilon_{\rm ph}}{1\,{\rm keV}}\right)^{-1} \;,
	\label{xxi}
\end{eqnarray}
where
\begin{eqnarray}
	\mu \equiv \left(\frac{\dot{M}}{5\times 10^{-5} 
		M_\odot\,{\rm yr}^{-1}}\right) \left(\frac{
		v_\infty}{2,000\,{\rm km}\,{\rm s}^{-1}}
		\right)^{-1} \;.
	\label{mu2}
\end{eqnarray}

Hence, for typical parameters we have $\Xi\ga 10^6$, which means that a tiny 
fraction of the radiation would be enough to fully ionize the gas in the 
wind. Then, the absorption opacity is negligible \citep{kro84}. The opacity 
in the wind to the X-ray photons is then dominated by the electron scattering 
opacity, $\kappa_\es = 0.2$ cm$^2$ g$^{-1}$, which is insensitive to 
the photon energy if the photon energy is much smaller than the electron 
mass energy \citep{akh65}. 

The X-ray optical depth in the wind is
\begin{eqnarray}
	\tau_\x \equiv \kappa_\es \int_r^\infty \rho dr = \iota\tau_w \;,
		\hspace{1cm} \iota \equiv \frac{\kappa_\es}{\kappa_w} \;.
	\label{taux}
\end{eqnarray}
The X-ray photospheric radius, defined by $\tau_\x=2/3$, is then
\begin{eqnarray}
	y_{\ph,\x} = \alpha \left[1-\left(1+\frac{2}{3\iota\tau_0}
		\right)^{1/(1-b)}\right]^{-1} \;.
	\label{yphx}
\end{eqnarray}

\section{Propagation of a Strong Shock Wave in the Wind}
\label{shock}

The propagation of a strong shock wave in a gas is determined by two competing
processes: the collection of mass from the ambient gas makes the shock wave
decelerate, and the steep downward gradient of the gas mass density makes the
shock wave accelerate. Based on previous self-similar analytical solutions 
and numerical works, \citet{mat99} have proposed a continuous and simple form 
for the shock velocity that accommodates both spherical deceleration and 
planar acceleration
\begin{eqnarray}
	v_s \propto \left(\frac{E_\in}{m}\right)^{1/2}
		\left(\frac{\rho r^3}{m}\right)^{-\beta_1} \;,
	\label{mm}
\end{eqnarray}
where $\beta_1\approx 0.2$. In the above equation, $E_\in$ is the explosion 
kinetic energy, $m(r)\equiv M(r) - M_\rem$, $M_\rem$ is the mass of the 
material that will become the supernova remnant, and $M(r)$ is the mass of the 
material contained in radius $r$.

After the shock has collected an enough amount of mass so that $m(r)$ does 
not change significantly any more, we have $v_s \propto\left(\rho r^3
\right)^{-\beta_1}$,
the behavior of the shock is purely determined by the profile of the mass 
density in the region that the shock is plowing into. Then, for a spherically 
symmetric gas, the shock wave accelerates when $d\left(\rho r^3\right)/dr <0$, 
and decelerates when $d\left(\rho r^3\right)/dr >0$. 

To generalize the formalism to the case of a relativistic shock wave, 
\citet{gna85} has suggested to replace the shock velocity $v_s$ on the 
left-hand side of equation~(\ref{mm}) by $\Gamma_s \beta_s$, where 
$\beta_s\equiv v_s/c$, $c$ is the speed of light, and $\Gamma_s \equiv 
\left(1-\beta_s^2\right)^{-1/2}$ is the Lorentz factor. The equation 
so obtained quite accurately describes both the limits of non-relativistic 
($\beta_s^2 \ll 1$, i.e., $\Gamma_s\approx 1$) and ultra-relativistic 
($\beta_s^2\approx 1$, i.e., $\Gamma_s\gg 1$) shocks. However, \citet{tan01} 
have shown that it is not accurate in the trans-relativistic regime 
($\beta_s^2$ close to $1$ but $\Gamma_s$ not large enough). \citet{tan01} 
have suggested the following formula for both non-relativistic and 
trans-relativistic shocks
\begin{eqnarray}
	\Gamma_s \beta_s &=& p\left(1+p^2\right)^{0.12} \;, \nonumber\\
	p &\equiv& 0.736 \left(\frac{E_\in}{m c^2}\right)^{1/2} 
		\left(\frac{\rho r^3}{m}\right)^{-0.187} \;.
	\label{tan}
\end{eqnarray}

With numerical simulation, \citet{tan01} have verified equation~(\ref{tan}) 
for trans-relativistic and accelerating shocks with $\Gamma_s \beta_s$ up to 
a value $\sim 10$. However, the limited numerical resolution in their 
code has not allowed them to follow the acceleration of a non-relativistic 
shock into the ultra-relativistic regime \citep{tan01}.

Although equation~(\ref{tan}) has never been tested on relativistic and
decelerating shock waves, in the non-relativistic limit it returns to the
formula of Matzner \& McKee, i.e. equation~(\ref{mm}), which applies to both
accelerating and decelerating shocks. Hence, we assume that 
equation~(\ref{tan}) applies to both accelerating and decelerating
relativistic shocks.
 
Because of the compactness of WR stars, the problem that we are solving
here is right in the trans-relativistic regime (as will be confirmed latter
in this paper). Thus we will use equation~(\ref{tan}) to calculate the 
momentum of a shock wave propagating in a wind of a WR star. In addition, 
since the wind contains a negligible amount of mass, at the radius where shock 
breakout takes place (either inside the star but close to its surface, or 
in the wind region), we have $m\approx M_\ej$, where $M_\ej$ is the ejected 
mass.

Although the equation for the shock movement that we use in this paper is
the same as that used by \citet{mat99} and \citet{tan01}, the mass density
profile in the wind of a star is very different from that in the interior
of a star. In the outer layer of a star the mass density drops quickly as 
the radius increases by a very small amount, as described by
equation~(\ref{rho_star}), Hence, as the shock wave approaches the surface
of the star, it always accelerates according to $v_s\propto \rho^{-\beta_1}$
since $m\approx M_\ej$ and $r\approx R_\star$. 

The situation is very different in a stellar wind. A shock wave propagating 
in a wind with a density given by 
equations~(\ref{rho_w}) and (\ref{vr}) accelerates in the region near the
stellar surface $r = R_\star$, but decelerates at large radius since $\rho
\propto r^{-2}$ and $d(\rho r^3)/dr >0$ for $r\gg R_\star$ (Fig.~\ref{slope}). 
The transition from acceleration to deceleration occurs at a radius determined 
by $d(\rho r^3)/dr = 0$, where the shock velocity reaches the maximum. The 
transition radius is found to be
\begin{eqnarray}
	R_a = (1+b) \alpha R_\star \;. \label{ra}
\end{eqnarray}

After passing the transition radius, the shock wave starts decelerating. The 
maximum $\Gamma_s \beta_s$ is then obtained by submitting $r=R_a$ into
equation~(\ref{tan})
\begin{eqnarray}
	\left(\Gamma_s \beta_s\right)_{\max}
		&=& p_{\max}\left(1+p_{\max}^2\right)^{0.12} \;,  
		\nonumber\\[1mm] 
	p_{\max} &=& 1.181\, [\alpha f(b)]^{-0.187} \nonumber\\
		&& \times \left(\frac{E_\in}
		{M_\ej c^2}\right)^{1/2} \left(\frac{\Psi}{M_\ej}
		\right)^{-0.187} \;,
	\label{pmax0}
\end{eqnarray}
where $f(b) \equiv (1+b)\left(1+1/b\right)^b$, and the mass function $\Psi$ 
is defined by
\begin{eqnarray}
	\Psi \equiv \frac{\dot{M} R_\star}{v_\infty} 
		= 1.654\times 10^{-9} M_\odot\, \mu  
		\left(\frac{R_\star}{3 R_\odot}\right) \;,
	\label{psi}
\end{eqnarray} 
where $\mu$ is defined by equation~(\ref{mu2}). 

The function $\Psi$ is an estimate of the mass contained in the photosphere
region of the wind. A correlation between $\Psi$ and $R_\ph$ is presented in 
Appendix~\ref{correlation}.

Submitting fiducial numbers in, we get
\begin{eqnarray}
	p_{\max} &=& 1.137\mu^{-0.187}\, \left[\frac{\alpha f(b)}{f(5)}
		\right]^{-0.187}\left(\frac{E_\in}{10^{52} {\rm ergs}}
		\right)^{0.5} \nonumber\\
		&&\times \left(\frac{M_\ej}{10 M_\odot}\right)^{-0.313}
		\left(\frac{R_\star}{3 R_\odot}\right)^{-0.187} \;.
	\label{pmax}
\end{eqnarray}
Thus, typically, the shock wave is trans-relativistic.

\section{Energy of the Radiation Contained in the Shock Front}
\label{energy}

The gas pressure behind a relativistic shock front, measured in the 
frame of the shocked gas, is \citep{bla76}
\begin{eqnarray}
	p_2 = (\gamma_2-1)(\hat{\gamma}_2\gamma_2+1)\rho c^2 \;,
\end{eqnarray}
where $\hat{\gamma}_2$ is the polytropic index of the shocked gas, $\gamma_2$ 
is the Lorentz factor of the shocked gas, and $\rho$ is the mass density of 
the unshocked gas. The Lorentz factor $\gamma_2$ is related to the Lorentz 
factor of the shock front $\Gamma = \Gamma_s$ by the equation~(5) of 
\citet{bla76}. Since WR winds are radiation-dominated, we have 
$\hat{\gamma}_2 = 4/3$. Then, we can approximate $p_2$ by
\begin{eqnarray}
	p_2 \approx F_p(\Gamma_s v_s)\, \rho \Gamma_s^2 v_s^2 \;,
\end{eqnarray}
where $F_p(\Gamma_s v_s)\sim 1$ is defined by
\begin{eqnarray}
	F_p(x) \equiv \frac{2}{3} + \frac{4}{21\left(1+0.4252\,
		x^2\right)^{0.4144}} \;,
	\label{fapprox}
\end{eqnarray}
which has the correct asymptotic values as $\Gamma_s v_s\rightarrow 
\infty$ (ultra-relativistic limit) and $\Gamma_s v_s\rightarrow 0$ 
(non-relativistic limit), and has a fractional error $<0.3\%$ in the 
trans-relativistic regime.

Denoting the temperature of the radiation behind the shock front by 
$T_2$, then the pressure of the radiation measured in the frame of the 
shocked gas is
\begin{eqnarray}
	\frac{1}{3} a T_2^4 \approx p_2 \approx F_p(\Gamma_s v_s)\,
		\rho \Gamma_s^2 v_s^2 \;,
	\label{p_r}
\end{eqnarray}
where $a$ is the radiation density constant.

A strong shock has a very narrow front. In the non-relativistic limit, the 
geometric thickness of the shock front is \citep{ims88,ims89}
\begin{eqnarray}
	\Delta r_s \approx \frac{c}{\kappa_\es\rho v_s} \;,  \label{ims}
\end{eqnarray}
where $\kappa_\es$ is the electron scattering opacity (see Sec.~\ref{wr}). 
That is, the thickness of the shock front is equal to the mean free path 
of photons multiplied by the optical depth of the shock
\begin{eqnarray}
	\tau_s = \frac{c}{v_s} \;.  \label{tau_s}
\end{eqnarray}

For an ultra-relativistic blast wave, the total energy stored in the shock
wave is proportional to $\Gamma_s^2 r^3$ and hence the thickness of the shell 
of shocked particles is $\sim r/\Gamma_s^2$ \citep{bla76}. That is, in the
ultra-relativistic limit, the thickness of the shock front measured in the 
rest frame is $\propto \Gamma_s^{-2}$. Hence, using the optical depth of the 
shock and that of the wind, we can estimate the geometric thickness of a
relativistic shock front in the rest frame by
\begin{eqnarray}
	\Delta r_s \approx \xi\frac{\tau_s}{\tau_\x}\frac{r}{\Gamma_s^2} \;,
	\label{ims_g}
\end{eqnarray}
where $\tau_\x$ is the X-ray optical depth (eq.~\ref{taux}), and
\begin{eqnarray}
	\xi &\equiv& \frac{\tau_\x}{\kappa_\es\rho r} 
		= \left|\frac{\partial\ln\tau_\x}{\partial\ln r}\right|^{-1}
		\nonumber\\
	        &=& \frac{y}{\alpha(b-1)}\left[1-\frac{\alpha}{y}-
		\left(1-\frac{\alpha}{y}\right)^b\right] \;.
	\label{xi}
\end{eqnarray}

Equation~(\ref{ims_g}) returns to equation~(\ref{ims}) in the non-relativistic
limit. The function $\xi(y)$ is an increasing function of $y$. As $y
\rightarrow\infty$, we have $\xi\rightarrow 1$. As $y\rightarrow \alpha$, we 
have $\xi\rightarrow 0$. (When $b=1$, $\xi$ is given by eq.~\ref{xi1} in 
Appendix~\ref{b1}.)

The total energy of the radiation contained in the shock front, measured 
in the frame of the shocked gas, is then
\begin{eqnarray}
	E_R \approx \frac{1}{3} \left(aT_2^4\right) 4\pi r^2 (\gamma_2
	        \Delta r_s) \approx \frac{4\pi\tau_s\gamma_2}{3\tau_\x 
		\Gamma_s^2} \xi \left(aT_2^4\right) r^3 \;,
	\label{e_r}
\end{eqnarray}
where the factor $1/3$ accounts for the fact that the energy density is
not uniform (concentrated at the boundary of the shock), and the factor 
$\gamma_2$ in $(\gamma_2 \Delta r_s)$ accounts for the Lorentz contraction. 
Submitting equation~(\ref{p_r}) into equation~(\ref{e_r}), we get
\begin{eqnarray}
	E_R \approx \frac{4\pi \tau_s\gamma_2}{\tau_\x \Gamma_s^2} \xi 
	        F_p(\Gamma_s v_s)\,\rho r^3 \Gamma_s^2 v_s^2 \;.
	\label{e_r2}
\end{eqnarray}

Using the definition of $\xi$, we have
\begin{eqnarray}
	E_R \sim \frac{4\pi\gamma_2c}{\Gamma_s^2 \kappa v_s} F_p r^2 
		\Gamma_s^2 v_s^2 \propto r^2 \Gamma_s v_s \;,
	\nonumber
\end{eqnarray}
since $F_p\sim 1$ and $\gamma_2/\Gamma_s \sim 1$.

In the accelerating region ($r<R_a$), $\Gamma_s v_s$ and $\gamma_2$ increase 
with $r$. Hence, the total energy contained in the shock front as measured
by the rest observer, $\gamma_2 E_R$, increases with $r$. 

In the decelerating region ($r>R_a$, $\rho\sim r^{-2}$), by 
equation~(\ref{tan}) we have, approximately, $\Gamma_s v_s \propto r^{-0.2}$, 
thus $E_R \propto r^{1.8}$. In the non-relativistic limit, $\gamma_2 E_R
\approx E_R \propto r^{1.8}$. In the 
ultra-relativistic limit, $\gamma_2 E_R\propto \gamma_2 r^{1.8} \propto 
r^{1.6}$ since $\gamma_s = \Gamma_s/\sqrt{2}\propto r^{-0.2}$. Hence, 
in the region of $r>R_a$, we also expect that the total energy contained in 
the shock front, $\gamma_2 E_R$, increases with $r$ although the shock is 
decelerating. This is caused by the fact that the volume contained in the 
shock front increases with $r$.

\section{Emergence of the Shock and the Characteristic Quantities}
\label{emergence}

Inside the star or deep inside the wind, because of the large optical depth 
in the gas photons have a diffusion velocity that is smaller than the 
velocity of the shock front, so that the radiation generated by the shock wave 
is trapped inside the boundary of the shock. As the shock wave moves toward 
the boundary of the photosphere, the optical depth in the gas drops, until a 
radius is reached where the diffusion velocity of photons begins to exceed 
the velocity of the shock front. Then, the radiation generated by the shock
wave starts to escape from the star to produce a bright flash, and the shock 
becomes visible to a remote observer.

Thus, the shock emerges at a radius where the optical depth of the gas to the
radiation generated by the shock is equal to the optical depth 
of the shock
\begin{eqnarray}
	\tau_\x = \frac{c}{v_s} \;,  \label{br_con}
\end{eqnarray}
since beyond that radius photons diffuse outward faster than the shock front 
moves \citep{ims88,ims89,mat99}. Since $v_s<c$ always, the shock must emerge 
at a radius where $\tau_\x >1$. By equations~(\ref{tau_r}) and (\ref{taux}),
the maximum breakout radius (determined by $\tau_\x=1$) is at
\begin{eqnarray}
	y_{\max} = \alpha\left[1-\left(1+\frac{1}{\iota\tau_0}
		\right)^{1/(1-b)}\right]^{-1} \;,
\end{eqnarray}
which is approached by an ultra-relativistic shock. [When $b=1$, the 
$y_{\max}$ is given by equation~(\ref{ymax1}).]

The evolution of $\Gamma_s \beta_s$ is determined by equation~(\ref{tan}), 
which can be recast into
\begin{eqnarray}
	\Gamma_s \beta_s &=& p\left(1+p^2\right)^{0.12} \;, \nonumber\\
	p &=& 1.181 \left(\frac{E_\in}{M_\ej c^2}\right)^{1/2}\left(
		\frac{\Psi}{M_\ej}\right)^{-0.187} \nonumber\\
		&& \times y^{-0.187}\left(1-\frac{\alpha}{y}
		\right)^{-0.187 b} \;,
	\label{tan2}
\end{eqnarray}
where equations~(\ref{rho_w})and (\ref{vr}) have been used, and $\Psi$ is
the mass function defined by equation~(\ref{psi}).

With equations~(\ref{tan2}), (\ref{taux}), and (\ref{tau_r}) (or \ref{tau1} 
if $b=1$), we can calculate the radius where the shock breakout takes place, 
$R_\br \equiv y_\br R_\star$, by numerically solving the algebraic 
equation~(\ref{br_con}).

After having $y_\br$, we can calculate the momentum of the shock wave at
$y=y_\br$, by equation~(\ref{tan2}).

\begin{figure*}
\vspace{2pt}
\includegraphics[angle=0,scale=0.75]{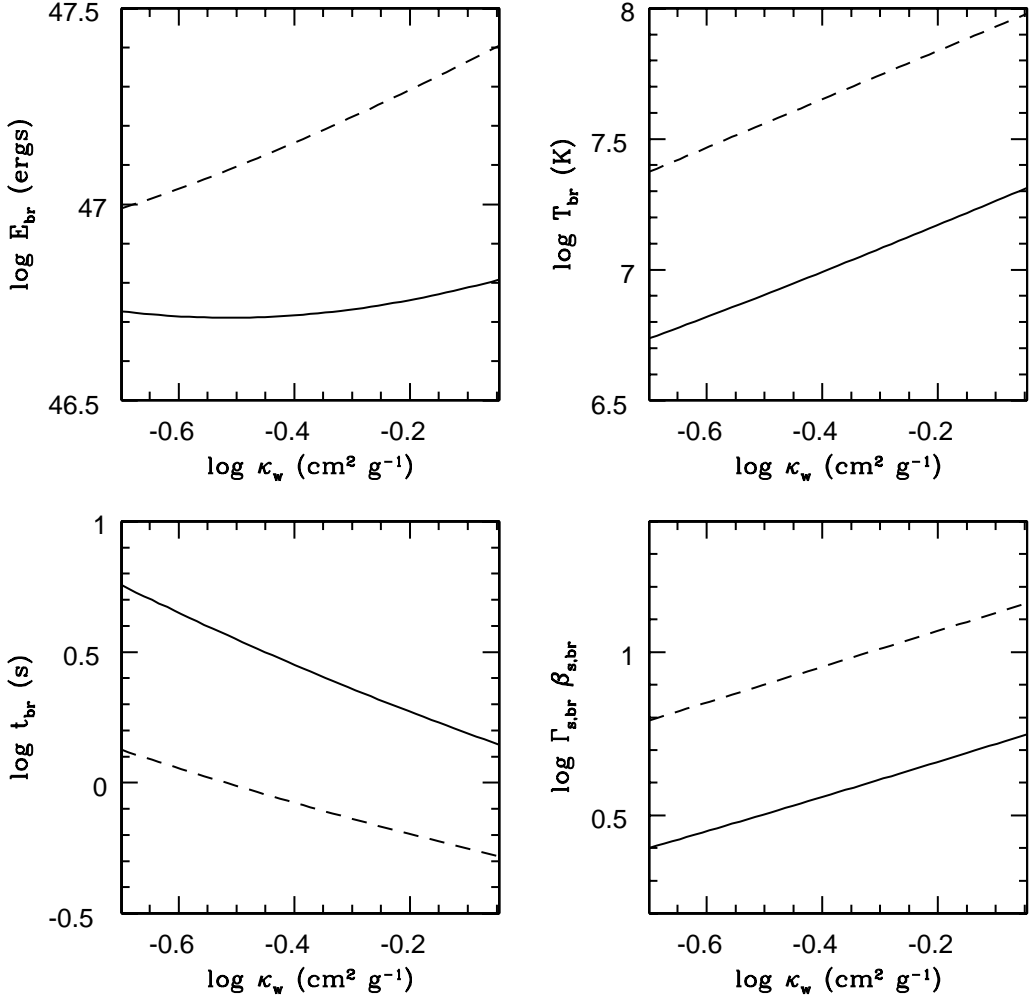}
\caption{Characteristic quantities of shock emergence as functions of 
the opacity in the stellar wind. The solid line corresponds to $\varepsilon
=10^{-2}$. The dashed line corresponds to $\varepsilon=10^{-3}$. Other 
parameters are: $b = 5$, $E_\in = 10^{52} {\rm ergs}$, $M_\ej = 10 
M_\odot$, and $R_\star = 3 R_\odot$.
}
\label{breakout_kappa}
\end{figure*}

Since the shock breakout occurs at a radius where $\tau_s = \tau_\x$ 
(eq.~\ref{br_con}), by equation~(\ref{e_r2}), the total energy of the 
radiation generated by the shock breakout as measured by a rest observer is
\begin{eqnarray}
	E_\br \equiv \left[\gamma_2 E_R\right]_{r=R_\br}
	        \approx \left.4\pi \xi F_\gamma^2 F_p\, \rho r^3
		\Gamma_s^2 v_s^2\right|_{r=R_\br} \;,
	\label{ebr0}
\end{eqnarray}
where $F_p = F_p(\Gamma_s v_s)\sim 1$, $F_\gamma = F_\gamma(\Gamma_s)\equiv
\gamma_2/\Gamma_s\sim 1$. The Lorentz factors $\gamma_2$ and $\Gamma_s$ are
related by the equation~(5) of \citet{bla76}. For the case of $\hat{\gamma}_2
=4/3$, $F_\gamma$ can be approximated by
\begin{eqnarray}
        F_\gamma(x) \approx \frac{1}{\sqrt{2}}+\frac{1-1/\sqrt{2}}
                {[1+0.9572\, (x - 1)]^{0.9325}} \;,
	\nonumber
\end{eqnarray}
which gives the correct asymptotic values at the non-relativistic limit 
($\Gamma_s\rightarrow 1$) and the ultra-relativistic limit ($\Gamma_s
\rightarrow \infty$), and has a fractional error $<0.08\%$ in the 
trans-relativistic case.

Submitting equations~(\ref{rho_w}) and (\ref{vr}) into equation~(\ref{ebr0}) 
and making use of equation~(\ref{psi}), we get
\begin{eqnarray}
	E_\br &\approx& \Psi c^2 \left[\xi F_\gamma^2 F_p \Gamma_s^2 
		\beta_s^2\right]_{y=y_\br} y_\br\left(1-\frac{\alpha}
		{y_\br}\right)^{-b} \nonumber\\
	&\approx& 1.48 \times 10^{46} {\rm ergs}\; \mu
		\left(\frac{R_\star}{3 R_\odot}\right) 
		\left(\frac{y_\br}{5}\right)
		\left(1-\frac{\alpha}{y_\br}\right)^{-b} \nonumber\\
		&&\times \left[\xi F_\gamma^2 F_p \Gamma_s^2 \beta_s^2
		\right]_{y=y_\br} \;.
	\label{ebr_eq}
\end{eqnarray}

\begin{figure*}
\vspace{2pt}
\includegraphics[angle=0,scale=0.75]{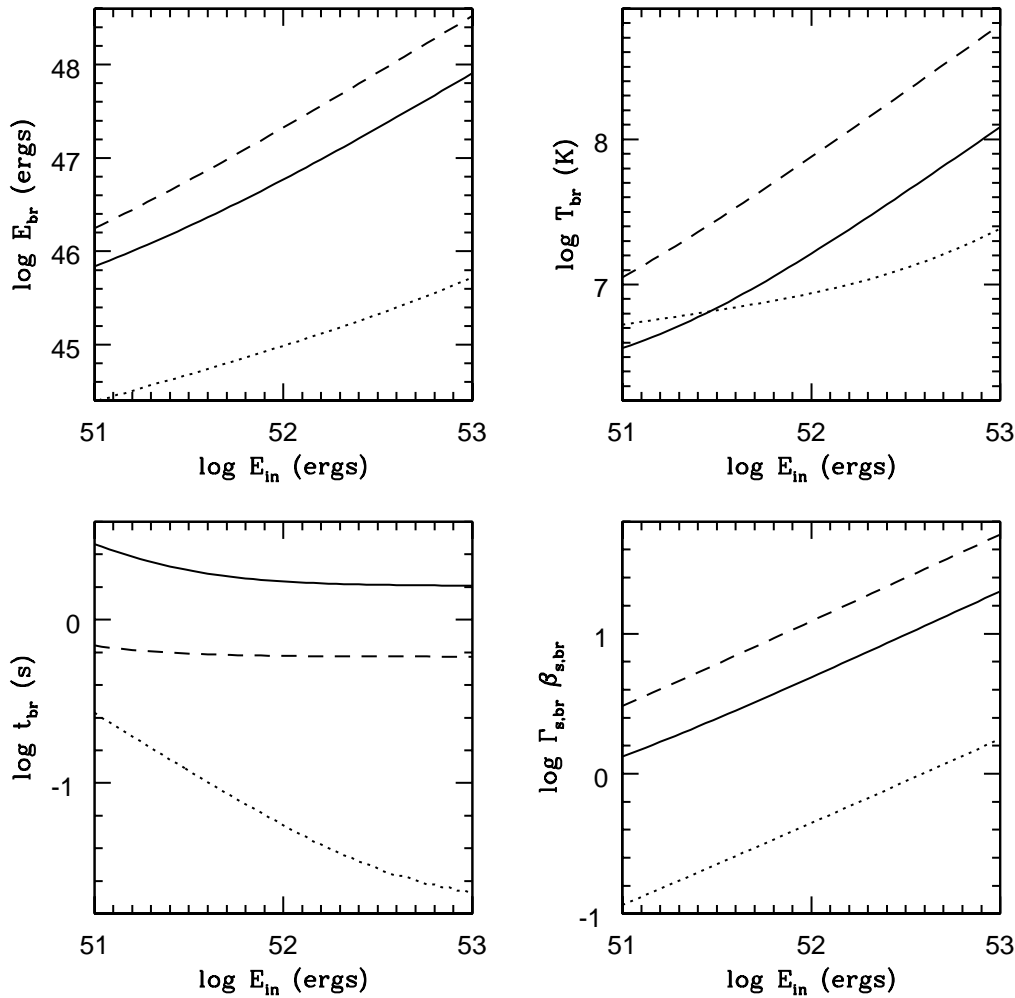}
\caption{Characteristic quantities of shock emergence as functions of 
the explosion kinetic energy. The solid line corresponds to $\varepsilon=
10^{-2}$. The dashed line corresponds to $\varepsilon=10^{-3}$. Other 
parameters are: $b = 5$, $\kappa_w = 0.7$ cm$^2$ g$^{-1}$, $M_\ej = 10 
M_\odot$, and $R_\star = 3 R_\odot$. The dotted line shows the solution 
for shock breakout from a star without a wind, with the same $M_\ej$, 
$R_\star$, and $\kappa_\star= 0.2$ cm$^2$ g$^{-1}$, $\zeta=1$.
}
\label{breakout_ein}
\end{figure*}

Similarly, from equation~(\ref{p_r}), we can obtain the temperature of the
radiation measured in a rest frame
\begin{eqnarray}
	T_\br &\equiv& \left[\gamma_2 T_2\right]_{r=R_\br} \nonumber\\
		&\approx& \left(\frac{3\Psi c^2}{4\pi a R_\star^3}
		\right)^{1/4} \left[F_\gamma F_p^{1/4}\Gamma_s^{3/2} 
		\beta_s^{1/2}\right]_{y=y_\br} \nonumber\\
		&&\times y_\br^{-1/2}\left(1-\frac{\alpha}{y_\br}
		\right)^{-b/4} \nonumber\\
	&\approx& 0.800\times 10^6 {\rm K}\; \mu^{0.25}
		\left(\frac{R_\star}{3 R_\odot}\right)^{-0.5}
		\left(\frac{y_\br}{5}\right)^{-0.5} \nonumber\\
		&&\times\left(1-\frac{\alpha}{y_\br}\right)^{-b/4}
		\left[F_\gamma F_p^{1/4}\Gamma_s^{3/2} \beta_s^{1/2}
		\right]_{y=y_\br} \;. 
	\label{tembr_eq}
\end{eqnarray}

The time-duration of the shock breakout event is set by the time spent by
a photon to diffuse out to the surface of the photosphere from the breakout
radius. Since at the radius of shock breakout the diffusion velocity of 
photons is equal to the velocity of the shock wave, we have
\begin{eqnarray}
	t_\br &\approx& \frac{R_{\ph,\x}-R_\br}{v_{s,\br}} = \frac{R_\star
		}{\beta_{s,\br} c} \left(y_{\ph,\x}-y_\br\right) \nonumber\\
		&\approx& 6.96\, {\rm s}\, \left(\frac{R_\star}{3 R_\odot}
		\right)\beta_{s,\br}^{-1} \left(y_{\ph,\x}-y_\br\right) \;,
	\label{tbr_eq}
\end{eqnarray}
where $R_{\ph,\x} = y_{\ph,\x} R_\star$ is the X-ray photospheric radius 
(eqs.~\ref{yphx} and \ref{yphx1}), and $v_{s,\br} = \beta_{s,\br} c 
\equiv v_s(r=R_\br)$ is the speed of the shock wave at the time of breakout.

\section{Results}
\label{result}

Unlike in the cases of non-relativistic and ultra-relativistic shock waves, 
where the quantities characterizing the transient event from the shock breakout
can be expressed with factorized scaling relations of input parameters (e.g., 
the eqs.~36--38 of Matzner \& McKee 1999), in the trans-relativistic case 
here we must numerically solve the relevant equations for the characteristic
quantities. 

\begin{figure*}
\vspace{2pt}
\includegraphics[angle=0,scale=0.75]{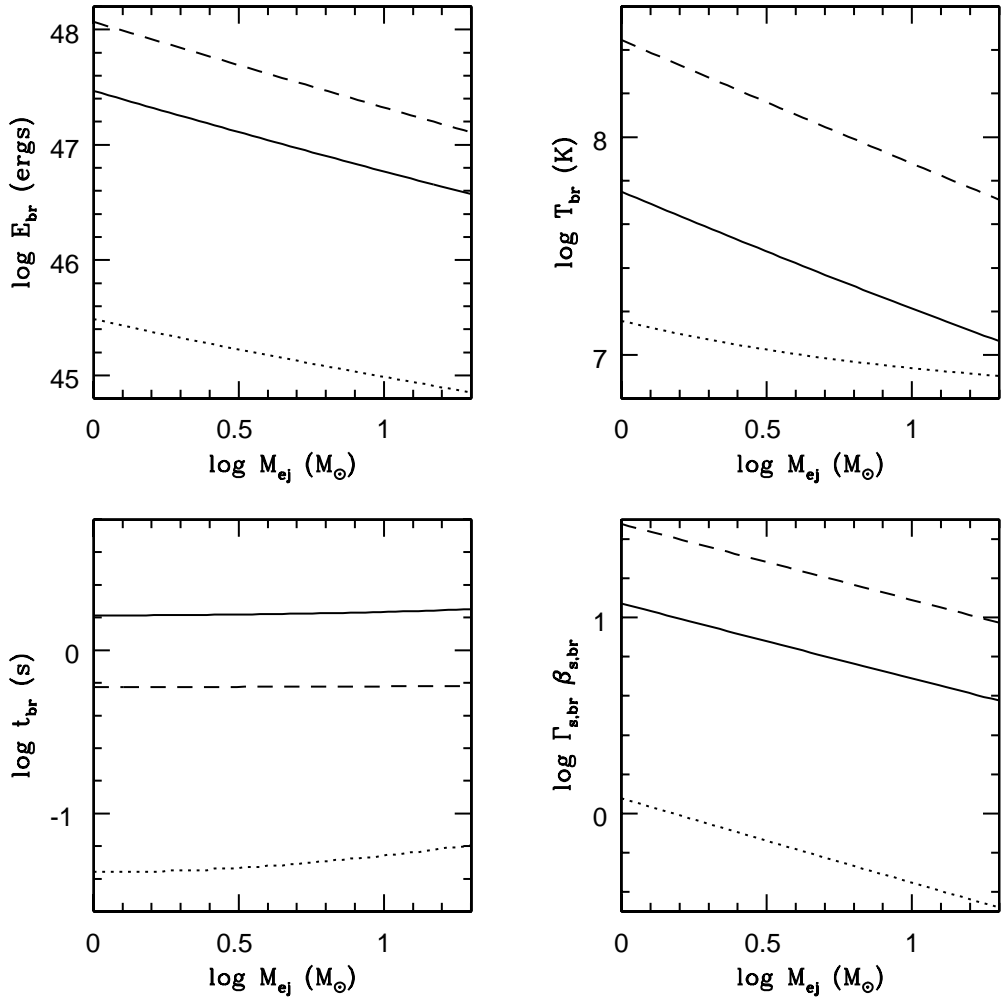}
\caption{Characteristic quantities of shock emergence as functions of 
the ejected mass. The solid line corresponds to $\varepsilon=10^{-2}$. 
The dashed line corresponds to $\varepsilon=10^{-3}$. Other parameters are: 
$b = 5$, $\kappa_w = 0.7$ cm$^2$ g$^{-1}$,  $E_\in = 10^{52}$ ergs, and 
$R_\star = 3 R_\odot$. The dotted line shows the solution for shock breakout 
from a star without a wind, with the same $E_\in$, $R_\star$, and 
$\kappa_\star= 0.2$ cm$^2$ g$^{-1}$, $\zeta=1$.
}
\label{breakout_mej}
\end{figure*}

All the relevant equations are in Sec.~\ref{emergence}, supplemented by the 
formulae for the wind mass function $\Psi$ and the optical depth in 
Secs.~\ref{wr} and \ref{shock} (and Appendix~\ref{b1} when $b=1$). We can 
eliminate $\dot{M}$ and $v_\infty$ from the equations by using
\begin{eqnarray}
	\Psi = \frac{80\pi (b-1) \alpha R_\star^2}{\kappa_w\left[(1-\alpha)
		^{1-b}-1\right]} \;,
	\label{psi_ka}
\end{eqnarray}
which is obtained by submitting equation~(\ref{tau0}) into the definition of
$\Psi$ (eq.~\ref{psi}). [When $b=1$, the corresponding equation is 
(\ref{psi1}).] Since $\alpha$ is a function of $\varepsilon$ and $b$ 
(eq.~\ref{alp}), we can then choose the input parameters to be $E_\in$, 
$M_\ej$, $R_\star$, $\varepsilon$, $b$, and $\kappa_w$.

Note, two opacities are involved in our model: $\kappa_w$, the optical opacity
in the wind of a WR star, which is used to calculate the mass density profile
in the wind before the shock wave passes through it; $\kappa_\x = \kappa_\es$,
the X-ray opacity in the wind, which is used to calculate the interaction of
the X-ray photons generated by the shock wave with particles in the wind
(see Sec.~\ref{wr}). Since $\kappa_\es = 0.2$ cm$^2$ g$^{-1}$ is a constant
but $\kappa_w$ is somewhat uncertain, we treat $\kappa_w$ as an input 
parameter.

Compared to the case of shock breakout from a star without a wind
\citep[and Appendix~\ref{star} in this paper]{mat99}, here we have two 
additional parameters: $\varepsilon$ and $b$, both describing the shape of 
the wind velocity profile (eqs.~\ref{vr} and \ref{alp}). However, in the case 
of a star, the opacity is fairly well determined so there are essentially 
only three parameters: the explosion energy $E_\in$, the ejected mass $M_\ej$, 
and the stellar radius $R_\star$. Although there is yet another parameter 
$\zeta \equiv \rho_1/\rho_\star$ (see Appendix~\ref{star}), which 
is typically $0.2$ for blue supergiants and $0.5$ for red supergiants 
\citep{cal04}, the characteristic quantities (at least the energy, the 
temperature, and the shock momentum) at shock breakout are very
insensitive to $\zeta$ \citep{mat99}. While for the problem here, i.e., a 
dense stellar wind surrounding a star, the opacity $\kappa_w$ is poorly known. 
Modeling of the WR winds indicates that $\kappa_w$ is in the range of 
$0.3$--$0.9$ cm$^2$ g$^{-1}$ at the sonic point, and slightly larger at 
larger radii \citep{nug02}. 

The parameter $\varepsilon$, which is the ratio of the wind velocity at the
stellar surface to the terminal velocity of the wind, is usually thought to 
be in the range of $0.001$--$0.1$, and typically around $0.01$ \citep{sch96}.

The parameter $b$, which characterizes the log-slope of the wind velocity in
the region near the stellar surface, is taken to be unity in the ``standard
model'' of stellar winds. However, as already mentioned in Sec.~\ref{wr}, for 
the winds of WR stars $b$ can be much larger than unity as argued by 
\citet{rob94}, \citet{sch97}, and \citet{lep99}, and is typically in the 
range of $4$--$6$ \citep{nug02}.

In our numerical modeling, we allow $\kappa_w$ to vary from $0.2$ to $0.9$
cm$^2$ g$^{-1}$, $\varepsilon$ to vary from $10^{-5}$ to $10^{-1}$, and
$b$ from 1 to 10. We allow $E_\in$ to vary from $10^{51}$ ergs (for normal
core-collapse supernovae) to $10^{53}$ ergs (for hypernovae), $M_\ej$ 
to vary from $1 M_\odot$ to $20 M_\odot$. Although WR stars are compact and
have small radii, 
to fully explore the effect of variation in the stellar radius on the 
results, we allow $R_\star$ to vary from $1 R_\odot$ to $30 R_\odot$. 
Whenever numbers are quoted, we refer to the fiducial values $\kappa_w= 0.7$ 
cm$^2$ g$^{-1}$, $\varepsilon = 0.01$, $b=5$, $E_\in = 10^{52}$ ergs, 
$M_\ej = 10 M_\odot$, and $R_\star = 3 R_\odot$, unless otherwise stated.

\begin{figure*}
\vspace{2pt}
\includegraphics[angle=0,scale=0.75]{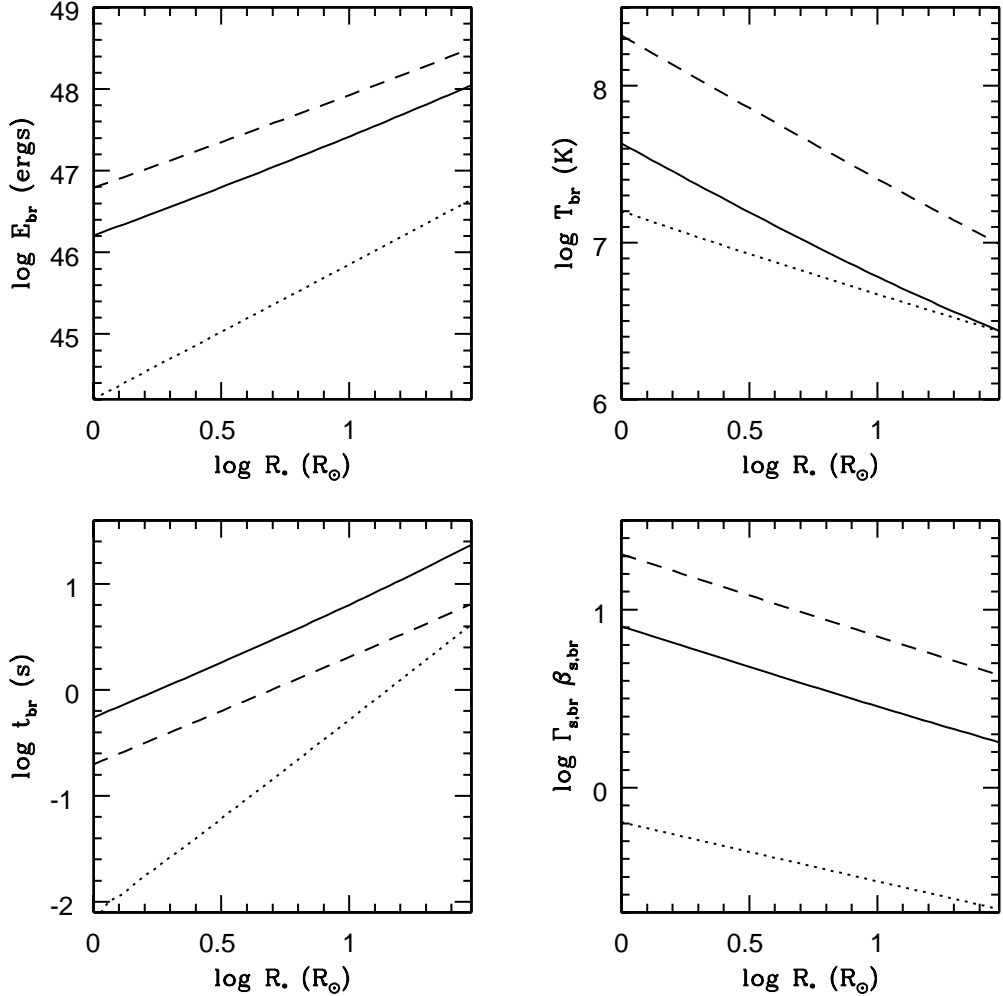}
\caption{Characteristic quantities of shock emergence as functions of 
the core radius of the star. The solid line corresponds to $\varepsilon=
10^{-2}$. The dashed line corresponds to $\varepsilon=10^{-3}$. Other 
parameters are: $b = 5$, $\kappa_w = 0.7$ cm$^2$ g$^{-1}$,  $E_\in = 10^{52}$ 
ergs, and $M_\ej = 10 M_\odot$. The dotted line shows the solution for 
shock breakout from a star without a wind, with the same $E_\in$, $M_\ej$, 
and $\kappa_\star = 0.2$ cm$^2$ g$^{-1}$, $\zeta=1$.
}
\label{breakout_rstar}
\end{figure*}

\begin{figure*}
\vspace{2pt}
\includegraphics[angle=0,scale=0.75]{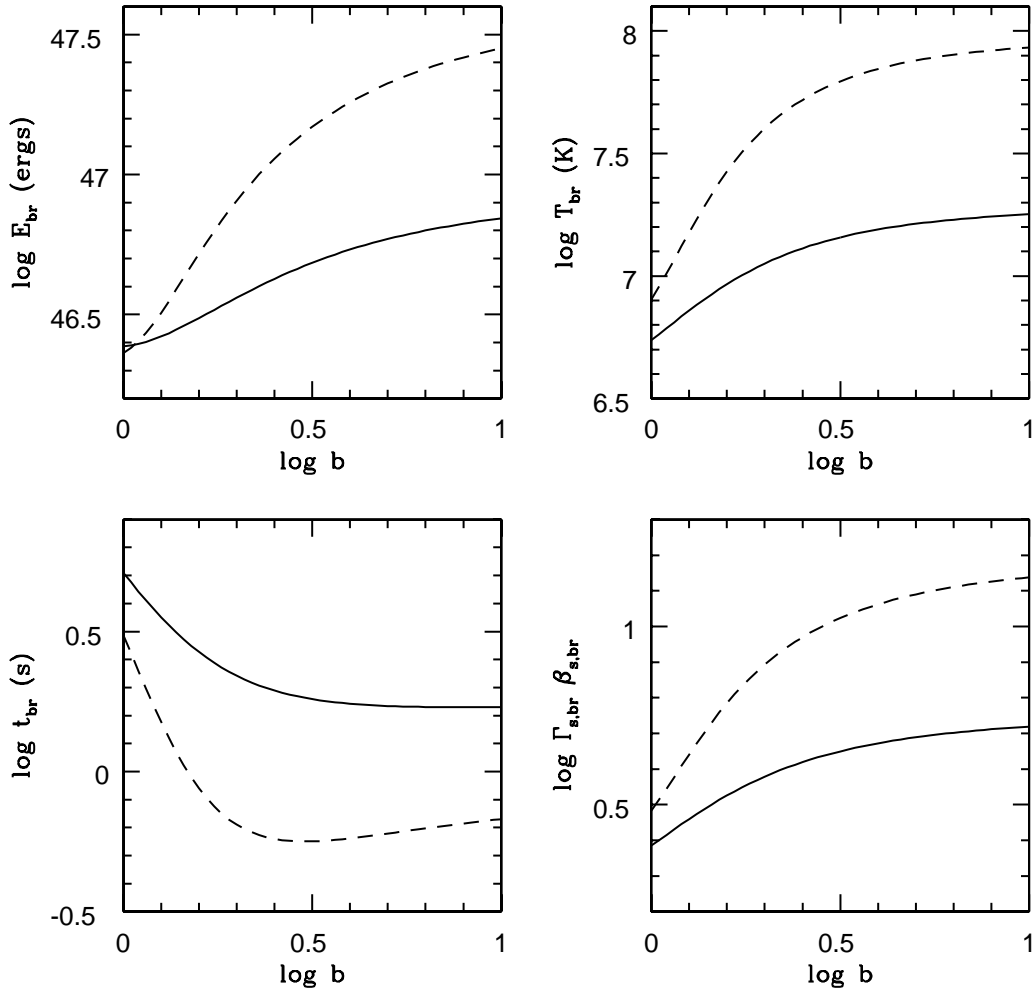}
\caption{Characteristic quantities of shock emergence as functions of 
the wind parameter $b$. The solid line corresponds to $\varepsilon=10^{-2}$. 
The dashed line corresponds to $\varepsilon=10^{-3}$. Other parameters 
are: $b = 5$, $\kappa_w = 0.7$ cm$^2$ g$^{-1}$, $E_\in = 10^{52}$ ergs, 
$M_\ej = 10 M_\odot$, and $R_\star = 3 R_\odot$.
}
\label{breakout_b}
\end{figure*}

Our numerical results for the characteristic quantities of the shock breakout,
including the total energy ($E_\br$, eq.~\ref{ebr_eq}), the temperature 
($T_\br$, eq.~\ref{tembr_eq}), the time-duration ($t_\br$, eq.~\ref{tbr_eq}), 
and the shock momentum ($\Gamma_{s,\br}\beta_{s,\br}$, eq.~\ref{tan2} with
$y=y_\br$) are presented in Figs.~\ref{breakout_kappa}--\ref{breakout_b}.

Figure~\ref{breakout_kappa} shows $E_\br$, $T_\br$, $t_\br$, and 
$\Gamma_{s,\br} \beta_{s,\br}$ as functions of the opacity $\kappa_w$.  
Solid lines correspond to $\varepsilon = 10^{-2}$. Dashed lines correspond 
to $\varepsilon =10^{-3}$. Other parameters take the fiducial values, as 
indicated in the figure caption. For $\varepsilon = 10^{-2}$, $E_\br$ is a
slow but not monotonic function of $\kappa_w$. For $\varepsilon = 10^{-3}$, 
$E_\br$ increases with $\kappa_w$. As $\kappa_w$ increases from $0.2$ 
to $0.9$ cm$^2$ g$^{-1}$, $E_\br$ increases by a factor $\approx 1.2$ when 
$\varepsilon =10^{-2}$, and $\approx 2.6$ when $\varepsilon =10^{-3}$. The 
temperature $T_\br$ increases by a factor $\approx 4$ in both cases. The 
opacity $\kappa_w$ has also an effect on $t_\br$, which decreases by a factor
when $\approx 4$ when $\varepsilon =10^{-2}$, and $\approx 2.6$ when 
$\varepsilon =10^{-3}$. Similar to the temperature, the momentum of the shock 
front is also an increasing function of $\kappa_w$. 
As $\kappa_w$ increases from $0.2$ to $0.9$ cm$^2$ g$^{-1}$, $\Gamma_{s,\br}
\beta_{s,\br}$ increases by a factor $\approx 2.2$ (for both $\varepsilon = 
10^{-2}$ and $\varepsilon = 10^{-3}$). Similar to the case of breakout from
a star, the results do not change dramatically with the opacity if $b$ is
around 5. Thus, the poor knowledge in the opacity in the stellar winds will 
not affect our results drastically.

All the dependence on $\kappa_w$ manifests itself though the mass function
$\Psi$ in equation~(\ref{psi_ka}) (and eq.~\ref{psi1} when $b=1$), which 
shows that $\Psi\propto \kappa_w^{-1}$. Then, from the condition for the 
shock breakout (eq.~\ref{br_con}), it can be checked that $y_\br$ decreases 
with $\kappa_w$, and $(1-\alpha/y_\br)^{-b}$ increases with $\kappa_w$.
From the dependence of $E_\br$, $T_\br$, $t_\br$, and $\Gamma_{s,\br} 
\beta_{s,\br}$ on $\Psi$ and $y_\br$, it is not hard to understand the trend 
shown in Fig.~\ref{breakout_kappa}. First, equation~(\ref{tan2}) implies
that $\Gamma_{s,\br} \beta_{s,\br}$ is a strong increasing function of 
$\kappa_w$. Then, equation~(\ref{tbr_eq}) implies that $t_\br$ is a 
decreasing function of $\kappa_w$. In equation~(\ref{ebr_eq}), $\Psi$ and 
$y_\br$ decrease with $\kappa_w$, but $\Gamma_{s,\br}^2 \beta_{s,\br}^2$
and $(1-\alpha/y_\br)^{-b}$ increase with $\kappa_w$. The overall result on
$E_\br$ is that shown in the top-left panel in Fig.~\ref{breakout_kappa}. 
Since the radius of shock breakout decreases with $\kappa_w$, the temperature
$T_\br$ must increase with $\kappa_w$.

Figure~\ref{breakout_ein} shows the same set of characteristic quantities 
as functions of the explosion kinetic energy. Symbols and values of parameters
are similar to Fig.~\ref{breakout_kappa} and are explained in the figure 
caption. To compare with the results for a star without a wind, we show with 
dotted lines the corresponding characteristic quantities calculated for the 
shock breakout from a star with the formulae in Appendix~\ref{star}, for the 
same values of $M_\ej$ and $R_\star$, and $\kappa_\star = 0.2$ cm$^2$ 
g$^{-1}$, $\zeta = 1$.

As the explosion energy increases from $10^{51}$ ergs to $10^{53}$ ergs, the 
breakout energy increases by a factor $\approx 117$ when $\varepsilon =
10^{-2}$, and $\approx 188$ when $\varepsilon = 10^{-3}$. This increasing 
rate is much faster than that in the case of breakout from a star, in which 
the breakout energy increases only by a factor of $\approx 22$. The increase 
in the temperature is also faster, which is by a factor of $\approx 33$ when 
$\varepsilon = 10^{-2}$, and $\approx 55$ when $\varepsilon = 10^{-3}$ for a 
star with a dense wind, and only $\approx 4.6$ for a star without a wind. 
While for the breakout time-duration, it appears that for the case of a 
stellar wind the time-duration does not change rapidly when $E_\in$ 
increases, in contrast to the case of a star. This is caused by the fact that 
when a star is surrounded by a dense stellar wind the shock wave has more
space for acceleration and hence at the time of emergence the shock front 
is more relativistic (see the panel for the shock momentum), its 
velocity approaches the speed of light limit. As we have seen in 
Sec.~\ref{emergence}, when the shock velocity approaches the speed of light, 
the breakout radius approaches $y_{\max}$. The distance between $y_{\max}$ 
and $y_\ph$ does not change with the explosion energy.

The momentum of the shock front varies with $E_\in$ at about the same rate 
for the case of a stellar wind and the case of a star. 

The curvature of the curves in Fig.~\ref{breakout_ein} confirms our claim 
at the beginning of this section that in the trans-relativistic case the 
characteristic quantities of shock breakout in general cannot be written 
as factorized scaling formulae of input parameters.

Figure~\ref{breakout_mej} shows the characteristic quantities as functions
of the ejected mass. As the ejected mass $M_\ej$
increases from $1 M_\odot$ to $20 M_\odot$, the breakout energy decreases by 
a factor $\approx 7.8$ when $\varepsilon =10^{-2}$, and $\approx 9.1$ when 
$\varepsilon = 10^{-3}$, faster than the case of a star for which the 
decreasing factor $\approx 4.4$. The temperature also drops faster. The 
variation in the breakout time-duration is not fast in both the cases of 
stellar winds and stars. That is, the time-duration of the shock breakout 
is not very sensitive to the ejected mass. The momentum of the shock front 
drops slightly slower than that in the case of a star.

Figure~\ref{breakout_rstar} shows the characteristic quantities as functions
of the core radius of the star, which, as in the case of a star \citep{mat99},
is the parameter that most dramatically affects the values of the 
characteristic quantities. As $R_\star$ increases from $1 R_\odot$ to $30 
R_\odot$, the breakout energy increases by a factor $\approx 69$ when 
$\varepsilon =10^{-2}$, and $\approx 51$ when $\varepsilon = 10^{-3}$. 
However, this factor is smaller than that in the case of star 
without a wind, which is $\approx 277$. The temperature drops very fast,
caused by the increase in the area of the surface emitting the radiation. As 
$R_\star$ increases from $1 R_\odot$ to $30 R_\odot$, the temperature drops 
by a factor of $\approx 16$ when $\varepsilon = 10^{-2}$, and $\approx 21$ 
when $\varepsilon = 10^{-3}$, in contrast to the factor $\approx 5.8$ in the 
case of a star. The variation in the stellar radius also has a dramatic effect 
on the breakout time-duration, although the effect is less prominent than
in the case of a star. The factor by which the breakout time-duration
increases is $\approx 43$ when $\varepsilon =10^{-2}$, and $\approx 32$ 
when $\varepsilon = 10^{-3}$, in contrast to that $\approx 590$ in the case
of a star. The momentum of the shock front drops by a factor $\approx 4.8$
when $\varepsilon =10^{-2}$, and $\approx 4.5$ when $\varepsilon = 10^{-3}$, 
similar to the factor $\approx 3$ in the case of a star.

\begin{figure*}
\vspace{2pt}
\includegraphics[angle=-90,scale=0.741]{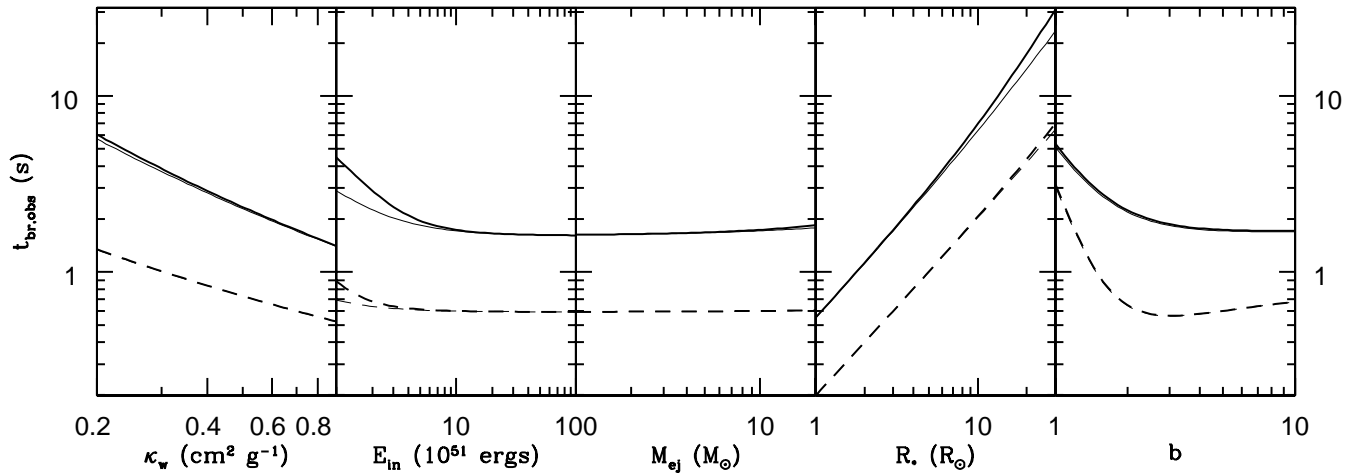}
\caption{The observed time-duration of shock breakout (defined by 
eq.~\ref{tbr_obs}) for the models in 
Figs.~\ref{breakout_kappa}--\ref{breakout_b} (from left to right). The solid 
line corresponds to $\varepsilon = 10^{-2}$. The dashed line corresponds to 
$\varepsilon = 10^{-3}$. The thin lines are the breakout time-duration 
$t_\br$ without the light-travel-time correction (eq.~\ref{tbr_eq}). The 
figure shows that in all models the light-travel-time correction is not 
important. In some models the thin and thick lines are even not visually 
distinguishable.
}
\label{t_obs}
\end{figure*}

Finally, in Fig.~\ref{breakout_b} we show the characteristic quantities as
functions of $b$. The breakout energy $E_\br$ increases with $b$. As $b$ 
increases from $1$ to $10$, $E_\br$ increases by a factor of $\approx 2.9$
when $\varepsilon =10^{-2}$, and $\approx 12$ when $\varepsilon = 10^{-3}$.
The smaller is $\varepsilon$, the faster does $E_\br$ increases with $b$.
However, if $b$ is in the range of 4--6, the change in $E_\br$ is not
essential, only by a factor of $\approx 1.2$--$1.3$. The temperature has a
similar trend. When $\varepsilon =10^{-2}$, the time-duration 
$t_\br$ decreases with $b$. When $\varepsilon =10^{-3}$, $t_\br$ decreases 
with $b$ until $b$ grows to a value $\approx 3$, beyond which $t_\br$ 
increases but slowly. The momentum of the shock front increases with $b$, 
caused by the fact that a larger $b$ results in a steeper density profile 
and an enhanced acceleration of the shock.

From Figs.~\ref{breakout_kappa}--\ref{breakout_rstar}, and 
Fig.~\ref{breakout_b} at $b=4$--$6$, the effects of variation in $\varepsilon$
from $10^{-2}$ to $10^{-3}$ can be summarized as follows: the breakout energy
$E_\br$ increases by a factor of $1.8$--$4$; the temperature $T_\br$ increases 
by a factor of $3$--$5$; the shock momentum $\Gamma_{s,\br} \beta_{s,\br}$
increases by a factor of $2.3$--$2.5$; and the time-duration $t_\br$ decreases
by a factor of $2.3$--$4.3$.

Figures~\ref{breakout_ein}--\ref{breakout_rstar} also show that, for a star 
with a dense wind the shock breakout is more energetic than that for a star 
without a wind. This is not surprising, since a star with a dense wind has 
effectively a larger radius so that the shock wave has more space and more 
time for acceleration. For the same set of common parameters 
($E_\in$, $M_\ej$, $R_\star$, but not the opacity), for typical parameters the 
total energy of the radiation from shock breakout is larger by a factor  
$>10$ if the star is surrounded by a dense wind. The momentum of the shock 
front is also larger by a factor $\sim 10$. The temperature does not show 
a universal trend because of increase in the shock breakout radius, but
generally it is larger if the progenitor is surrounded by a dense wind
due to the great enhancement in the breakout energy. The time-duration is 
larger for the case of stellar winds as an obvious result of increase in 
the effective radius of the star. 

The shock breakout occurs inside the maximum acceleration radius $R_a$ 
(eq.~\ref{ra}) in all the models presented in 
Figs.~\ref{breakout_kappa}--\ref{breakout_b}.

In the above calculations of the breakout time-duration, the light-travel-time 
has not been taken into account. In other words, $t_\br$ is the duration 
measured in the supernova frame. The duration observed by a remote observer, 
$t_{\br,\obs}$, differs from $t_\br$ by an effect caused by the travel-time of 
light---which arises from the fact that an observer will see more distant 
annuli of the stellar disk with a time-delay \citep{ens92}. The effect of
light-travel-time could be extremely important when the $t_\br$ calculated by 
equation~(\ref{tbr_eq}) is short, which is definitely true here since WR 
stars are compact. We approximate the observed time-duration of the shock 
breakout event by
\begin{eqnarray}
	t_{\br,\obs} = \sqrt{t_\br^2+t_\l^2} \;,
	\label{tbr_obs}
\end{eqnarray}
where $t_\l$ is the light-travel-time.

In the calculation of the light-travel-time $t_\l$, the relativistic beaming
effect must be taken into account since the shock wave in our models is 
relativistic \citep{kat94}. In the ultra-relativistic limit, the beaming 
angle is $\theta\sim 1/\Gamma_{\ph,X}$, where $\Gamma_{\ph,X}$ is the Lorentz 
factor of the photosphere which can estimated by $\Gamma_{\ph,X}\approx 
\Gamma_s$. In the non-relativistic limit, we should have $\theta=\pi/2$. 
Hence, we use an interpolation formula for $\theta$
\begin{eqnarray}
	\theta = \frac{\pi}{\pi(\Gamma_s-1)+2} \;. \label{theta}
\end{eqnarray}
Then, the light-travel-time is
\begin{eqnarray}
	t_\l = \frac{R_{\ph,\x}}{c}(1-\cos\theta) \;.
\end{eqnarray}
When $\Gamma_s\gg 1$, we have $t_\l \approx R_{\ph,\x}/(2\Gamma_s^2 c)$.

\begin{table*}
\centering
\begin{minipage}{160mm}
\caption{Models of Type Ibc supernova explosion and the predicted 
characteristic parameters for the shock breakout. Input parameters: $E_\in$, 
$M_\ej$, $R_\star$, $\kappa_w$, $\varepsilon$, and $b$. Output parameters: 
$y_{\ph,\x}$, $y_\br$, $(\Gamma_s\beta_s)_{\br} = \Gamma_{s,\br}\beta_{s,
\br}$, $E_\br$, $T_\br$, $t_{\br,\obs}$, and $\mu$. Models 1--4 are normal 
SNe Ibc. Models 5--7 are hypernovae. 
}
\label{model}
\begin{tabular}{llllllllllllll}
\hline
Model\hspace{0.cm} & ${E_\in}^{\rm a}$\hspace{0.cm} & 
${M_\ej}^{\rm b}$\hspace{0.cm} & ${R_\star}^{\rm c}$\hspace{0.cm} & 
${\kappa_w}^{\rm d}$\hspace{0.08cm} & ${\varepsilon}^{\rm e}$\hspace{0.42cm} & 
${b}^{\rm f}$\hspace{0.08cm} & ${y_{\ph,\x}}^{\rm g}$\hspace{0.cm} & 
${y_\br}^{\rm h}$\hspace{0.27cm} & ${(\Gamma_s\beta_s)_{\br}}^{\rm i}$ & 
${E_\br}^{\rm j}$\hspace{0.19cm} & ${T_\br}^{\rm k}$\hspace{0.cm} & 
${t_{\br,\obs}}^{\rm l}$\hspace{0.cm} & ${\mu}^{\rm m}$\\
\hline
1 & 1 & 3 & 3 & 0.7 & 0.01 & 5 & 1.73 & 1.45 & 1.98 & 1.3 & 5.4 & 2.8 & 0.30\\
2 & 1 & 4 & 3 & 0.2 & 0.02 & 5 & 4.24 & 2.45 & 0.818 & 1.2 & 1.9 & 25 & 1.7\\
3 & 1.5 & 6 & 5 & 0.5 & 0.01 & 1 & 3.11 & 1.61 & 0.760 & 1.3 & 1.7 & 35 & 2.4\\
4 & 2 & 2 & 5 & 0.7 & 0.001 & 5 & 1.31 & 1.22 & 6.72 & 19 & 28 & 1.0 & 0.095\\
5 & 40 & 10 & 3 & 0.2 & 0.01 & 5 & 3.21 & 2.53 & 5.73 & 22 & 16 & 4.9 & 1.0\\
6 & 50 & 10 & 10 & 0.7 & 0.01 & 5 & 1.73 & 1.50 & 7.52 & 140 & 22 & 5.5 & 0.98\\
7 & 60 & 15 & 10 & 0.7 & 0.002 & 5 & 1.39 & 1.28 & 13.7 & 320 & 62 & 2.6 & 0.32\\
\hline
\end{tabular}
$^{\rm a}$Explosion kinetic energy in units of $10^{51}$ ergs.\\
$^{\rm b}$Ejected mass in units of $M_\odot$.\\
$^{\rm c}$Core radius of the progenitor (the radius at the optical depth of 
20), in units of $R_\odot$.\\
$^{\rm d}$Optical opacity in the wind, in units of cm$^2$ g$^{-1}$.\\
$^{\rm e}$Ratio of the wind velocity at the stellar surface (where $r=
R_\star$) to the terminal velocity of the wind (eq.~\ref{alp}).\\
$^{\rm f}$Parameter specifying the profile of the wind velocity 
(eq.~\ref{vr}).\\
$^{\rm g}$Radius of the X-ray photosphere in units of $R_\star$.\\
$^{\rm h}$Radius of the shock front at the time of shock breakout in units of 
$R_\star$.\\
$^{\rm i}$Momentum of the shock front (eq.~\ref{tan2}) at the time of shock 
breakout.\\
$^{\rm j}$Total energy of the radiation from the shock breakout, in units 
of $10^{46}$ ergs.\\
$^{\rm k}$Temperature of the radiation from the shock breakout, in units of 
$10^6\, {\rm K} = 0.08617$ keV.\\
$^{\rm l}$Observed time-duration of the shock breakout event, in units of 
seconds.\\
$^{\rm m}$Observable defined by the mass-loss rate and the terminal velocity 
of the stellar wind through eq.~(\ref{mu2}).\\
\end{minipage}
\end{table*}

For the models presented in Figs.~\ref{breakout_kappa}--\ref{breakout_b}, we
have calculated the light-travel-time correction to the observed time-duration
of the shock breakout event. The results are shown in Fig.~\ref{t_obs}. It
turns out that the light-travel-time correction is not important. This is
caused by the fact that for relativistic shock breakout the light-travel-time
is significantly reduced by the the relativistic beaming effect.

Numerical results for a set of supernova and WR star models are presented in 
Table~\ref{model}. From these results we find that the efficiency of 
converting the supernova explosion energy to the shock breakout energy, 
defined by the ratio of the breakout energy to the explosion energy, is 
typically in the range of $10^{-4}$--$10^{-5}$. This efficiency is smaller 
than that in the case of Type II supernovae, which is typically $\sim 
10^{-3}$ if the progenitor 
is a red supergiant, or $\sim 10^{-4}$ if the progenitor is a blue supergiant.
This is again caused by the fact that WR stars have much smaller radii than
red and blue supergiants.

\section{Application to GRB 060218/SN 2006aj}
\label{application}

\begin{figure}
\vspace{2pt}
\includegraphics[angle=0,scale=0.463]{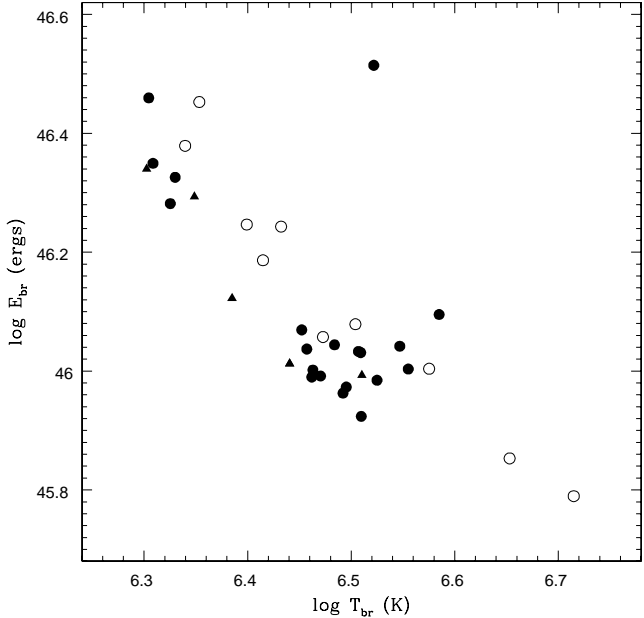}
\caption{Energy versus temperature of shock breakout in Type Ibc supernovae
produced by core-collapse of a sample of WR stars with $R_\ph/R_\star >2$.
}
\label{ebr_tem_wrs}
\end{figure}

\begin{figure}
\vspace{2pt}
\includegraphics[angle=0,scale=0.47]{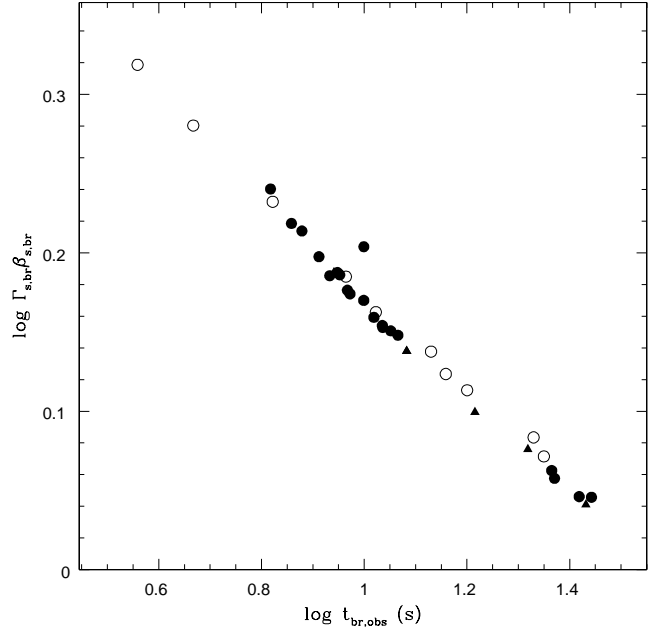}
\caption{Shock momentum versus the observed time-duration of shock breakout
in Type Ibc supernovae produced by core-collapse of the same sample of WRs 
in Fig.~\ref{ebr_tem_wrs}.
}
\label{gsbr_tbr_wrs2}
\end{figure}

As stated in the Introduction, recently it has been claimed that supernova 
shock breakout has been observed in the early X-ray emission of GRB 060218, 
based on the observation that a fraction ($\approx 20\%$) of the radiation 
in the lightcurve (from 159~s up to $\sim 10,000$~s after the trigger of the 
burst) is a soft black-body of temperature $\approx 0.17$ keV \citep{cam06}. 
The total energy estimated for this black-body component is $\approx 10^{49}$ 
ergs in the $0.3$--$10$ keV band, and $\approx 2\times 10^{49}$ in bolometric 
(S. Campana, private communication). A reanalysis carried out by \citet{but06} 
revealed an even larger energy in the black-body, which is $\approx 2 \times 
10^{50}$ ergs, with a duration $\approx 300$~s. 

The overall constraint on the black-body component in the early X-ray 
afterglow of GRB 060218 is summarized as follows: the total energy 
$\ga 10^{49}$ ergs, the temperature is in the range of $0.1$--$0.19$ keV 
(i.e., $1.2$--$2.2\times 10^6$ K), and the duration $\ga 300$~s 
\citep{cam06,but06}.

In this Section, we apply the procedure developed in previous sections to
calculate the characteristic quantities of the shock breakout event for 
SN 2006aj with the assumption that the supernova was produced by the 
core-collapse of a WR star surrounded by a dense wind, and examine if the 
black-body component in GRB 060218 can be interpreted by the shock 
breakout in SN 2006aj.

\begin{figure}
\vspace{2pt}
\includegraphics[angle=0,scale=0.476]{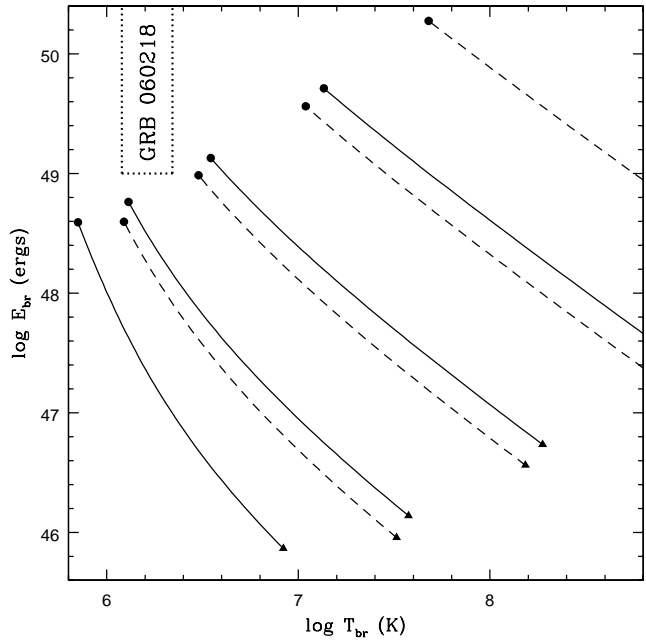}
\caption{The breakout energy versus the breakout temperature (with $b=5$). 
Solid lines correspond to $\kappa_w=0.2$ cm$^2$ g$^{-1}$. Dashed lines
correspond to $\kappa_w=0.9$ cm$^2$ g$^{-1}$. Different solid lines (and 
different dashed lines) correspond to different values of $\varepsilon$: 
$10^{-2}$, $10^{-3}$, $10^{-4}$ and $10^{-5}$ (upward). Along each line the
stellar radius $R_\star$ varies from $1 R_\odot$ (the triangle) to $100
R_\odot$ (the point). The supernova explosion energy $E_\in = 2\times 10^{51} 
{\rm ergs}$. The ejected mass $M_\ej = 2 M_\odot$. The region bounded by the 
dotted line indicates the observational constraint on the total energy and 
the temperature of the black-body component in the X-ray afterglow of 
GRB 060218.
}
\label{tembr_ebr}
\end{figure}

\begin{figure}
\vspace{2pt}
\includegraphics[angle=0,scale=0.476]{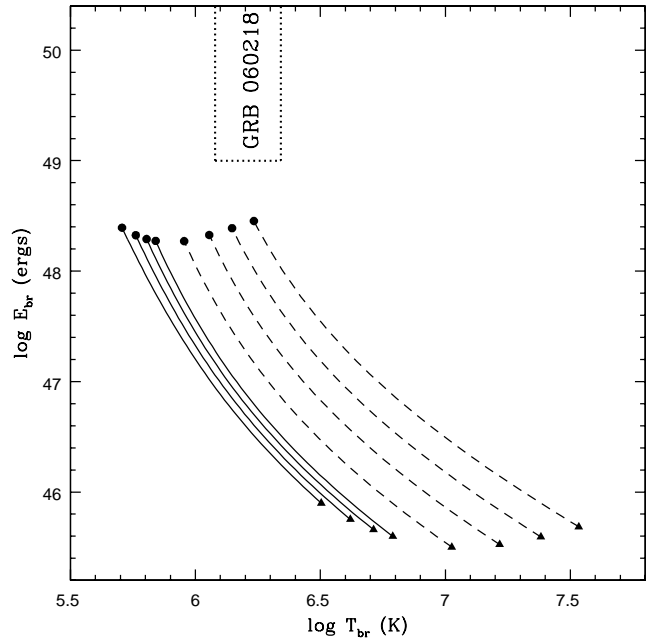}
\caption{Same as Fig.~\ref{tembr_ebr} but with $b=1$.
}
\label{tembr_ebr2}
\end{figure}

First, we apply the procedure to the WR stars in Fig.~\ref{rph}, which are
among the best studied catalog of WR stars with model-determined
stellar and photospheric radii \citep{ham95,koe95,gra98}. We pick up only 
stars with $y_\ph = R_\ph/R_\star>2$, since otherwise the $\varepsilon$ 
given by our simplified model would be too small (Fig.~\ref{alpha}). The 
sample so selected consists of total 36 stars, including 20 Galactic 
WCs, 10 Galactic WNs, and 6 LMC WCs. The majority of WC stars have been 
included. Since theses stars were modeled by the standard stellar wind model
(Appendix~\ref{b1}), we choose $b=1$. The $\alpha$ parameter is then obtained 
from the published values of $y_\ph=R_\ph/R_\star$ in the papers cited above
by equations~(\ref{tau0_1}) and (\ref{yph1}). The value of $\varepsilon$ is 
calculated with equation~(\ref{alp}). Then, with the published values of 
$\dot{M}$, $v_\infty$, and $R_\star$, the constant mean opacity $\kappa_w$ 
can be calculated with equation~(\ref{psi1}).

For the supernova explosion energy and the ejected mass, we take the values 
obtained by modeling the spectra and the lightcurve of SN 2006aj: 
$E_\in = 2\times 10^{51}$ ergs, and $M_\ej = 2 M_\odot$ \citep{maz06a}. 
Then, for each star we have all of the six parameters needed for calculating 
the characteristic quantities of shock breakout.

Our results are shown in Fig.~\ref{ebr_tem_wrs} for the breakout energy
versus the breakout temperature, and in Fig.~\ref{gsbr_tbr_wrs2} for
the breakout shock momentum versus the breakout time-duration. Note, here 
the breakout time-duration has included the light-travel-time 
(eq.~\ref{tbr_obs}), so it corresponds to the observed time-duration. From 
Fig.~\ref{ebr_tem_wrs} we see that, although the temperature is in the range 
of the black-body component in GRB 060218 for several WC stars (on the left 
end), the total energy of the radiation arising from the shock breakout never 
exceeds $10^{47}$ ergs, i.e., always smaller than the total energy of the 
observed black-body component in GRB 060218 by more than two orders of 
magnitude.

From Fig.~\ref{gsbr_tbr_wrs2}, the time-duration of the shock breakout never 
exceeds 100 s, also well below the observational limit on the black-body 
component in GRB 060218.

Thus, it appears that none of the stars in the considered sample of WRs is
able to produce a supernova with shock breakout energy that is large enough 
to explain the black-body component observed in the early X-ray afterglow of 
GRB 060218.

Of course, GRBs are rare events compared to supernovae \citep{pod04}, they
may require progenitors that are in more extreme conditions than the WRs in 
our sample. To test the possibility for explaining the black-body component 
in GRB 060218 with shock breakout from WR stars in a larger parameter space, 
we have calculated a number of models with a large range of parameters.
The results are shown in Fig.~\ref{tembr_ebr} ($b=5$) and 
Fig.~\ref{tembr_ebr2} ($b=1$). The explosion energy and the ejected mass are 
fixed at $E_\in = 2\times 10^{51}$ ergs and $M_\ej = 2 M_\odot$, as obtained 
by modeling the spectra and the lightcurve of SN 2006aj \citep{maz06a}. We 
allow
$\varepsilon$ to vary from $10^{-2}$ to $10^{-5}$. For the opacity $\kappa_w$,
we choose two extreme values: $0.2$ cm$^2$ g$^{-1}$ (solid lines) and $0.9$ 
cm$^2$ g$^{-1}$ (dashed lines). The observational bound on the total energy 
and the temperature of the black-body component in the early X-ray emission 
of GRB 060218 is shown in the figures by the region bounded by the dotted 
lines.

The radius of the star, $R_\star$, which is the parameter that the 
characteristic quantities of the shock breakout are most sensitive to, is 
allowed to vary from $1 R_\odot$ to $100 R_\odot$, covering a space of
radii that is more than enough for WR stars.

Figures~\ref{tembr_ebr} and \ref{tembr_ebr2} show that to explain the
black-body component observed in the early X-ray emission of GRB 060218, the
radius of the progenitor WR star must be $\ga 100 R_\odot$. It is very 
unlikely that there exist WR stars having so large stellar radii. Although
it is possible to get $E_\br > 10^{49}$ ergs with $R_\star<100 R_\odot$ if
$\varepsilon$ is very small and/or $\kappa_w$ is very large, the corresponding
$T_\br$ would be too high to be consistent with the temperature of the
black-body component in GRB 060218.

\section{Summary, Conclusions, and Discussions}
\label{sum}

We have presented a simple model for calculating the characteristic
quantities (total energy, temperature, time-duration, and shock momentum) 
for the flashes arising from shock breakout in Type Ibc supernovae produced 
by the core-collapse of Wolf-Rayet stars surrounded by dense stellar
winds. The wind velocity is modeled by equation~(\ref{vr}), a profile that 
is often adopted in the study of stellar winds. However, in contrast to the 
case for O-stars where the parameter $b$ is close to unity, for WR star
winds $b$ can be much larger and is usually in the range of $4$--$6$ 
\citep{nug02}. The opacity in the wind, $\kappa_w$, is assumed to be a 
constant, which is a reasonable approximation for the calculation of the 
optical depth since the opacity varies with radius very slowly compared to 
the mass density of the wind \citep{nug02}. Modeling of the opacity in the 
winds of WR stars indicates that $\kappa_w$ is in the range of $0.3$--$0.9$ 
cm$^2$ g$^{-1}$ \citep{nug02}.

Our model is an extension of the existing model for calculating the 
characteristic quantities for supernova shock breakout from a star without 
a wind, which is suitable for Type II supernovae 
\citep{ims88,ims89,mat99,tan01}. Due to the compactness of WR stars, the 
shock momentum is expected to be trans-relativistic at the time of breakout. 
Thus, we have followed \citet{bla76} and \citet{tan01} to take into account 
the relativistic effects. 

Because of the large optical depth in the wind, the supernova shock breakout 
occurs in the wind region rather than in the interior of the star. This is 
equivalent to say that the presence of a dense stellar wind effectively 
increases the radius of the star. As a result, the shock has more space and 
more time for acceleration, and the shock breakout appears to be more energetic
than in the case for the same star but the effect of the stellar winds is 
not taken into account (see. e.g., Blinnikov et al. 2002). 

The formulae for determining the radius where the shock breakout occurs
and that for computing the characteristic quantities for the radiation 
arising from the shock breakout are collected in Sec.~\ref{emergence}. They
include equations~(\ref{br_con}), determining the breakout radius; 
(\ref{tan2}), evaluating the momentum of the shock;
(\ref{ebr_eq}), (\ref{tembr_eq}), and (\ref{tbr_eq}), calculating the
energy, temperature, and the time-duration of the radiation from shock
breakout. Although exact and analytic solutions are impossible because of
the trans-relativistic nature of the problem, all the equations are
algebraic and a simple numerical program is able to calculate all the
characteristic quantities. The model contains six input parameters:
the explosion kinetic energy ($E_\in$), the ejected mass ($M_\ej$),
the core radius of the star ($R_\star$, the radius where the
optical depth $\tau_w=20$), the opacity in the wind ($\kappa_w$), the
parameter $b$ specifying the wind velocity profile, and the ratio of
the wind velocity at the stellar surface (where $r=R_\star$) to the terminal 
velocity of the wind ($\varepsilon$).

Our numerical results are summarized in 
Figs.~\ref{breakout_kappa}--\ref{breakout_b} and Table~\ref{model}.  
Figs.~\ref{breakout_kappa}--\ref{breakout_b} illustrate how the
characteristic quantities vary with the input parameters. As in the case
of shock breakout from a star without a wind, the core radius of the star is
the most important parameter affecting the results. That is, the
characteristic quantities are most sensitive to the variation in the
stellar radius. This feature leads to the possibility for distinguishing
the progenitors of supernovae by observing the flashes from the shock 
breakout \citep{cal04}. In addition, in the case of dense stellar winds, 
the results are more sensitive to the variation in the supernova explosion 
kinetic 
energy. For example, roughly speaking, $E_\br \propto E_\in$ when the star 
has a dense wind, in contrast to $E_\br \propto E_\in^{0.6}$ in the case of 
a star without a wind. Overall, the shock breakout from a star with a dense 
wind is more energetic than that from a star without a wind.
For a star of the same radius, and for the same explosion kinetic energy 
and ejected mass, the total energy released by the shock breakout is larger 
by a factor $> 10$ if the star is surrounded by a thick wind. The 
time-duration is also larger, and the shock momentum at the time of
breakout is more relativistic.

For explosion energy $E_\in=10^{51}$ ergs, ejected mass $M_\ej=3 M_\odot$,
and stellar radius $R_\star = 3 R_\odot$ (typical values for normal SNe Ibc), 
we get breakout energy $E_\br \approx 1.3\times 10^{46}$ ergs, 
temperature $T_\br \approx 5.4\times 10^6$ K $\approx 0.46$ keV, and
observed time-duration $t_{\br,\obs}\approx 2.8$ s if other parameters take 
fiducial values ($\kappa = 0.7$ cm$^2$ g$^{-1}$, $b=5$, and $\varepsilon = 
0.01$). For $E_\in=5\times 10^{52}$ ergs, $M_\ej=10 M_\odot$, and $R_\star 
= 10 R_\odot$ (typical values for hypernovae), we get $E_\br \approx 
1.4\times 10^{48}$ ergs, $T_\br \approx 2.2\times10^7$ K $ \approx 1.9$ keV, 
and $t_{\br,\obs}\approx 5.5$ s. More numerical results are shown in 
Table~\ref{model}.

We have applied our model to GRB 060218/SN 2006aj, in which a soft
black-body component has been observed in the early X-ray emission of the 
GRB and has been interpreted as an evidence for the supernova shock breakout 
\citep{cam06}. We take the values of the supernova explosion energy and the 
ejected mass obtained by modeling the spectra and the lightcurve of the 
supernova \citep{maz06a}. We find that, the energy released by the supernova 
shock breakout in a thick wind
of a WR progenitor star is generally too small to explain the black-body 
radiation in GRB 060218. To obtain the breakout energy and the temperature 
that are consistent with the observational constraint, the core radius of 
the progenitor WR star has to be $> 100 R_\odot$, which is much too large 
for a WR star. Thus, we conclude that the black-body component in the X-ray 
afterglow of GRB 060218 cannot be interpreted by the shock breakout in the 
underlying supernova. Instead, it must originate from other processes which 
might be related to the GRB outflow (see, e.g., Fan, Piran \& Xu 2006). 
This conclusion is in agreement with the analysis by Ghisellini, Ghirlanda 
\& Tavecchio (2006).

One may argue that GRB-connected supernovae should be highly aspherical so 
that our spherical model might have under-estimated the energy of the shock 
breakout. The effect of explosion asymmetry can be estimated as follow.
Assume that the explosion produces a shock wave in a solid angle $\Omega\equiv 
4\pi\omega <4\pi$ with a kinetic explosion energy $E_\in$, which ejects a mass
$M_\ej$ from the progenitor. The shock wave is symmetric in the azimuthal 
direction and does not expand to the outside of $\Omega$. The motion of the 
shock wave would then be the same as that of a spherical shock wave 
($\omega=1$) with a kinetic explosion energy $\omega^{-1} E_\in$ and an 
ejected mass $\omega^{-1} M_\ej$, assuming that the progenitor is spherically 
symmetric. Then, by equation~(\ref{tan2}), $p \propto E_\in^{1/2} 
M_\ej^{-0.313} \omega^{-0.187} = \left(\omega^{-0.374} E_\in\right)^{1/2} 
M_\ej^{-0.313}$. That is, the motion of the asymmetric shock wave can be 
calculated by equation~(\ref{tan2}) but with $E_\in$ replaced by a larger 
$E_\in^\prime = \omega^{-0.374} E_\in$. Then, by Fig.~\ref{breakout_ein}, 
the temperature $T_\br$, the shock momentum $\Gamma_{s,\br} \beta_{s,\br}$, 
and the {\em isotropic-equivalent} energy $E_\br$ of the asymmetric shock 
breakout are larger than that in a spherical explosion with the same $E_\in$ 
and $M_\ej$. However, the time-duration $t_\br$ is not sensitive to $\omega$.

Indeed, aspherical explosion has
been claimed to be observed in the luminous Type Ic SN 2003jd, in which 
the double-lined profiles in the nebular lines of neutral oxygen and 
magnesium revealed in later-time observations by Subaru and Keck are
explained as results of observing an aspherical supernovae along a 
direction almost perpendicular to the axis of the explosion \citep{maz05}. 
However, for SN 2006aj, there is no any evidence for aspherical explosion. 
Observation on the radio afterglow and modeling of it indicate that the 
outflow associated with GRB 060218 is mildly relativistic so should be more 
or less spherical (Soderberg et al. 2006; Fan, Piran \& Xu 2006, see also
Li 2006).

We should also remark that whether the progenitors of GRBs are surrounded by 
dense winds is still an open question. Although a wind-type density 
profile is naturally expected for the environment surrounding a GRB as
its progenitor is broadly thought to be a massive star, observations on the
GRB afterglows have revealed that most of the afterglow data are consistent
with a constant density external medium and only a handful of bursts can
be well modeled by the wind model 
\citep[and references therein]{ber03,zha04,pan05,fry06}. For the case of 
GRB 060218, modeling of its radio afterglow also does not favor a dense 
circum-burst wind profile \citep{sod06,fan06}.

A theoretical argument against strong winds surrounding GRB progenitors
comes from the consideration of angular momentum 
\citep[and references therein]{yoo05,woo06a}. For a black hole formed from
the core-collapse of a massive star to have a disk rotating around it and
to launch a relativistic jet, the progenitor star must rotate rapidly with the
specific angular momentum in the core $j\ga 3\times 10^{16}$ cm$^2$ s$^{-1}$
\citep{mac99}. To satisfy this requirement, the progenitor star should not
have had a phase with an intense stellar wind since a dense wind is very 
effective in removing angular momentum. Given the fact that the mass-loss 
rate of a star sensitively depends on its metallicity \citep{vin05} and the 
observations that GRBs prefer to occur in galaxies with low metallicity
\citep{fyn03,hjo03a,lef03,sol05,fru06,sta06}, it is reasonable to expect that 
the progenitors of GRBs should 
not have dense stellar winds surrounding them. Even in this situation, 
however, the radius of the massive progenitor star is also very unlikely to 
be large enough ($>100 R_\odot$) to explain the black-body component in 
GRB 060218 since its progenitor star has only a mass $\sim 20 M_\odot$ as
obtained by modeling the supernova lightcurve and spectra \citep{maz06a}. 
In addition, 
if the progenitor does not have a thick wind, then in calculating the results 
for the shock breakout one should use the formulae in Appendix~\ref{star} 
for a star without a wind. But in Sec.~\ref{result} we have seen that the 
formulae for a star without a wind lead to smaller total energy in the 
radiation from the shock breakout than the formulae for a star with a dense 
wind.

In spite of the disappointing result on GRB 060218/SN 2006aj, our model is 
expected to have important applications to Type Ibc supernovae since whose 
progenitors are broadly believed to be WR stars. In addition, some Type II 
supernovae appear also to be related to progenitor stars with intensive
stellar winds, e.g. SNe IIn (also called IIdw) \citep{ham04}. Observations 
on the transient events from supernova shock breakout will be the most 
powerful approach for diagnosing the progenitors of supernovae. For this goal 
we would like to mention {\it LOBSTER}, an upcoming space observatory 
dedicated to detect soft X-ray flashes from shock breakout in supernovae 
\citep{cal04}.

\section*{Acknowledgments}

The author thanks S. Campana for useful communications and sharing data,
and B. Paczy\'nski for many inspiring discussions on gamma-ray bursts, 
supernovae, and shock breakout. He also thanks an anonymous referee for
a very helpful report which has led significant improvements to the paper.

\appendix

\section{Optical Depth in a Wind in the Standard Model}
\label{b1}

In the standard model of stellar winds the parameter $b$ in 
equation~(\ref{vr}) is assumed to be unity. Then, the integral in 
equation~(\ref{tau}) gives
\begin{eqnarray}
	\tau_w = \tau_0 \ln \left(1-\frac{\alpha}{y}\right)^{-1} \;,
	\label{tau1}
\end{eqnarray}
where
\begin{eqnarray}
	\tau_0 \equiv \frac{A}{\alpha R_\star} = \frac{20}{ 
		\ln \left(1-\alpha\right)^{-1}} \;.
	\label{tau0_1}
\end{eqnarray}

The ratio of the photospheric radius (at $\tau_w = 2/3$) to the stellar core
radius (at $\tau_w = 20$) is
\begin{eqnarray}
	y_\ph = \frac{\alpha}{1-\exp[-2/(3\tau_0)]} \;,
	\label{yph1}
\end{eqnarray}
which approaches 1 as $\alpha\rightarrow 1$, and 30 as $\alpha\rightarrow
0$.

The X-ray photospheric radius is at
\begin{eqnarray}
	y_{\ph,\x} = \frac{\alpha}{1-\exp[-2/(3\iota\tau_0)]} \;.
	\label{yphx1}
\end{eqnarray}

The corresponding mass function $\Psi$ (eq.~\ref{psi}), when $\dot{M}$ and 
$v_\infty$ are eliminated (by using eq.~\ref{tau0_1}), is
\begin{eqnarray}
	\Psi = \frac{80\pi\alpha R_\star^2}{\kappa_w \ln(1-\alpha)^{-1}} \;.
	\label{psi1}
\end{eqnarray}

The parameter $\xi = \left|\partial\ln\tau_\x/\partial\ln r\right|^{-1}$ 
(Sec.~\ref{energy}) is
\begin{eqnarray}
	\xi = \frac{y}{\alpha}\left(1-\frac{\alpha}{y}\right) \ln \left(1-
		\frac{\alpha}{y}\right)^{-1} \;,
	\label{xi1}
\end{eqnarray}
which approaches unity as $y\rightarrow\infty$, and approaches zero as 
$y\rightarrow \alpha$.

The maximum radius where the shock breakout occurs (see Sec.~\ref{emergence})
is given by
\begin{eqnarray}
	y_{\max} = \frac{\alpha}{1-\exp(-1/\iota\tau_0)} \;.
	\label{ymax1}
\end{eqnarray}

\section{Shock Breakout from a Star without a Wind}
\label{star}

The mass density in an outer layer of a star is described by a power law
(see, e.g., Matzner \& McKee 1999)
\begin{eqnarray}
	\rho = \rho_1 x^n \;, \label{rho_star}
\end{eqnarray}
where $x\equiv 1-r/R_\star$, $n$ is related to the polytropic index 
$\hat{\gamma}$ by $\hat{\gamma} = 1+1/n$. When $\hat{\gamma} = 4/3$, we
have $n=3$.

The optical depth in the star is
\begin{eqnarray}
	\tau_\star = \tau_0 x^{n+1}\;, \hspace{1cm}
	\tau_0 \equiv \frac{\kappa_\star \rho_1 R_\star}{n+1} \;,
\end{eqnarray}
where $\kappa_\star$ is the opacity.

Near the stellar surface we have $r\approx R_\star$, so the shock  
accelerates according to equation~(\ref{tan}) with $m\approx M_\ej$
and $r\approx R_\star$ \citep{tan01}.

The geometric thickness of the shock front is
\begin{eqnarray}
	\Delta r_s \approx \frac{\tau_s}{\Gamma_s^2\kappa_\star\rho}
	        = \xi \frac{\tau_s}{\tau_\star} \frac{R_\star x}{\Gamma_s^2}
		\;,
\end{eqnarray}
where $\xi = 1/(n+1)$, $\tau_s = c/v_s$.

The shock breakout occurs at a radius where $\tau_\star = \tau_s$. The 
minimum value of $x_\br$, which occurs when $v_s\rightarrow c$, is 
\begin{eqnarray}
	x_{\min} = \tau_0^{-1/(n+1)} \;,
\end{eqnarray}
corresponding to the maximum breakout radius $r_{\max} = R_\star(1 
-x_{\min})$.

The pressure of the gas behind the shock front, measured in the
frame of the shocked gas, is still given by equation~(\ref{p_r}),
from which the temperature of the shock emergence can be calculated.

The total energy of radiation in the shock emergence, measured in the
rest frame, is
\begin{eqnarray}
	E_\br \approx \left.4\pi \xi F_\gamma^2 F_p\, \rho R_\star^3
		(\Gamma_s v_s)^2 x \right|_{r=R_\br} \;.
\end{eqnarray}

The time-duration of the shock breakout is
\begin{eqnarray}
	t_\br \approx \frac{R_\star x_\br}{v_{s,\br}} \;.
\end{eqnarray}

The input parameters include $E_\in$, $M_\ej$, $R_\star$, $\kappa_\star$,
and $\zeta\equiv \rho_1/\rho_\star$, where $\rho_\star\equiv M_\ej/
R_\star^3$.

\section{A Correlation in Wolf-Rayet Star Parameters}
\label{correlation}

\begin{figure}
\vspace{2pt}
\includegraphics[angle=0,scale=0.469]{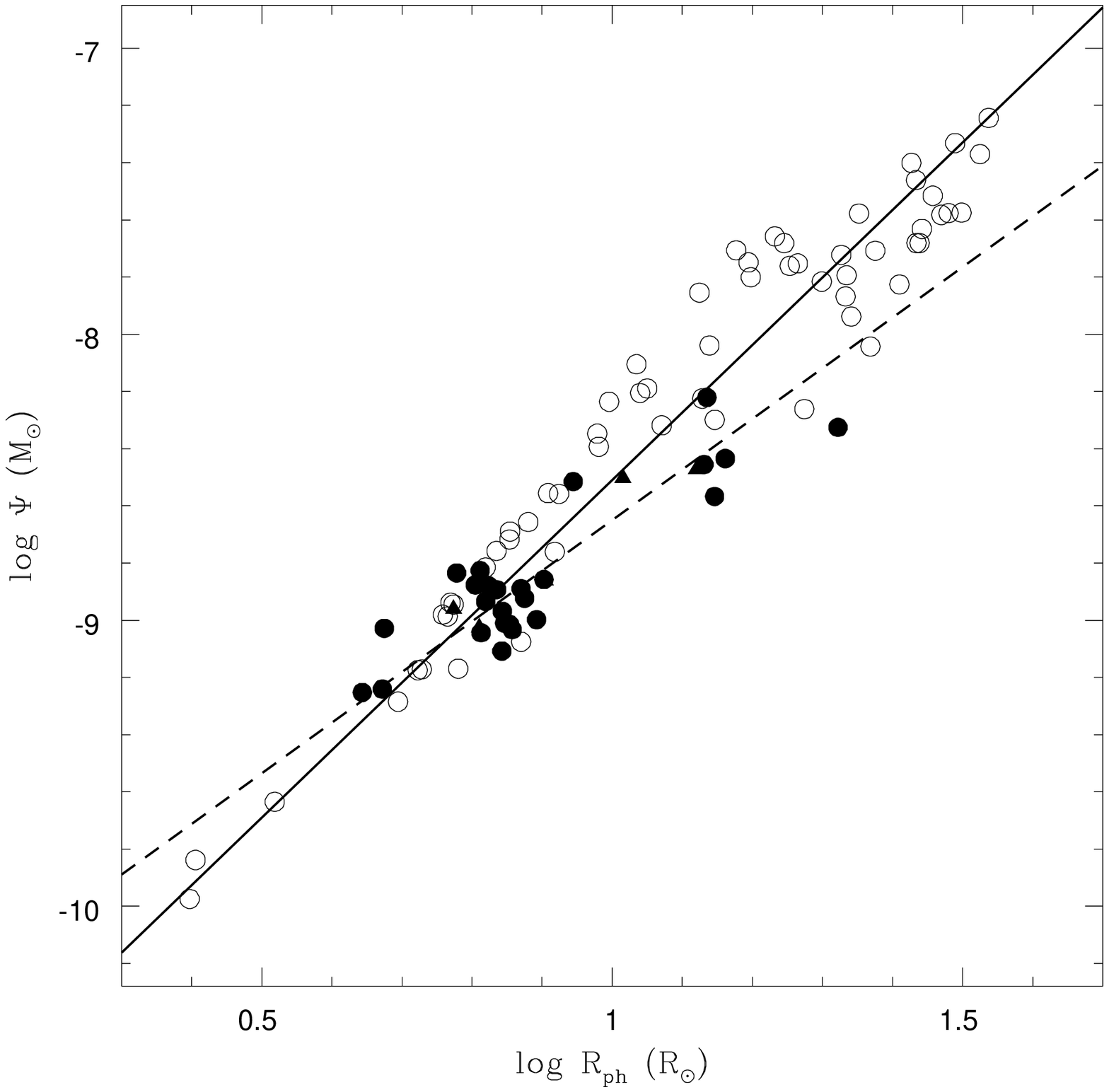}
\caption{The mass function $\Psi$ defined by eq.~(\ref{psi}) against the 
photospheric radius for the sample of WRs in Fig.~\ref{rph}. Clearly
there is a strong correlation between $\Psi$ and $R_\ph$. The solid straight 
line is the best fit to all the data by eq.~(\ref{psi_r}). The dashed 
straight line is the best fit to the WC stars (filled symbols) by 
eq.~(\ref{psi_r2}).
}
\label{mvr}
\end{figure}

From the parameters of the 92 Galactic and LMC WR stars presented in 
Fig.~\ref{rph}, a correlation between $\Psi=\dot{M} R_\star/v_\infty$ 
(eq.~\ref{psi}) and $R_\ph$ can be derived. 

In Fig.~\ref{mvr}, we plot $\log \Psi$ against $\log R_\ph$ for the 92
WRs. Clearly, there is a strong correlation between $\Psi$ and $R_\ph$.
The relation is best fitted by
\begin{eqnarray}
	\log \Psi = -10.87 + 2.36\, \log R_\ph
	\label{psi_r}
\end{eqnarray}
for all stars, and
\begin{eqnarray}
	\log \Psi = -10.42 + 1.77\, \log R_\ph
	\label{psi_r2}
\end{eqnarray}
for WC stars only, where $\Psi$ is in units of $M_\odot$, and $R_\ph$ is
in units of $R_\odot$.

To the knowledge of the author the relation does not exist in the literature
so is presented here, although it is irrelevant to the subject of the paper.

\bsp

\label{lastpage}

\end{document}